\documentclass[aps,prx,twocolumn,superscriptaddress]{revtex4-1}

\usepackage{graphicx}
\usepackage[caption=false]{subfig}
\usepackage{multirow}

\usepackage{nopageno, tikz, amsmath, amsfonts, mathrsfs,bm}

\newcommand{\fq}[1]{\Phi_\mathrm{ext}={#1}\Phi_0}
\newcommand{\fqn}{\Phi_\mathrm{ext}}
\newcommand{\fqo}{\Phi_1}
\newcommand{\fqt}{\Phi_2}
\newcommand{\g}{|g\rangle}
\newcommand{\e}{|e\rangle}

\begin{document}

\title{A fluxonium-based artificial molecule with a tunable magnetic moment}

\author{A.~Kou}
\affiliation{Departments of Applied Physics and Physics, Yale University, New Haven, CT 06520, USA}
\author{W.~C.~Smith}
\affiliation{Departments of Applied Physics and Physics, Yale University, New Haven, CT 06520, USA}
\author{U.~Vool}
\affiliation{Departments of Applied Physics and Physics, Yale University, New Haven, CT 06520, USA}
\author{R.~T.~Brierley}
\affiliation{Departments of Applied Physics and Physics, Yale University, New Haven, CT 06520, USA}
\author{H.~Meier}
\affiliation{Departments of Applied Physics and Physics, Yale University, New Haven, CT 06520, USA}
\author{L.~Frunzio}
\affiliation{Departments of Applied Physics and Physics, Yale University, New Haven, CT 06520, USA}
\author{S.~M.~Girvin}
\affiliation{Departments of Applied Physics and Physics, Yale University, New Haven, CT 06520, USA}
\author{L.~I.~Glazman}
\affiliation{Departments of Applied Physics and Physics, Yale University, New Haven, CT 06520, USA}
\author{M.~H.~Devoret}
\affiliation{Departments of Applied Physics and Physics, Yale University, New Haven, CT 06520, USA}
%\email[]{Your e-mail address}
%\homepage[]{Your web page}
%\thanks{}
%\altaffiliation{}

\date{\today}

\begin{abstract}
Engineered quantum systems allow us to observe phenomena that are not easily accessible naturally. The LEGO\textsuperscript{\textregistered}-like nature of superconducting circuits makes them particularly suited for building and coupling artificial atoms. Here, we introduce an artificial molecule, composed of two strongly coupled fluxonium atoms, which possesses a tunable magnetic moment. Using an applied external flux, one can tune the molecule between two regimes: one in which the ground-excited state manifold has a magnetic dipole moment and one in which the ground-excited state manifold has only a magnetic quadrupole moment. By varying the applied external flux, we find the coherence of the molecule to be limited by local flux noise.  The ability to engineer and control artificial molecules paves the way for building more complex circuits for protected qubits and quantum simulation.
\end{abstract}

%\pacs{}
%\keywords{}

\maketitle

\section{Introduction}
Superconducting circuits are a flexible platform for building artificial atoms. By choosing the proper combination of inductors, capacitors, and Josephson junctions, the experimentalist can tailor superconducting circuits to be governed by a variety of Hamiltonians and be insensitive to certain decoherence mechanisms \cite{Nakamura1999,VanderWalCH1999,Vion2002,Paik2011,Barends2013}. A particular example is the fluxonium atom, which consists of a small Josephson junction in parallel with a superinductance \cite{Manucharyan2009,Masluk2012}. Its spectrum maintains the anharmonic structure associated with single Cooper-pair phenomena while suppressing the detrimental effect of offset charges. At the external flux sweet spot, its ground-excited state manifold, $\{\g,\e\}$, is rendered flux-noise insensitive to first order.

Given the success of building artificial atoms using superconducting circuits, one may wonder whether we can build novel \textit{artificial molecules}. These circuits would have degrees of freedom that are delocalized between constituent artificial atoms, in analogy with the electronic motion that is delocalized between several nuclei in natural molecules. Here, the experimentalist would be able to tailor the type and strength of coupling between individual artificial atoms. In particular, by cleverly choosing the coupling between constituent atoms, the experimentalist could make the states of an artificial molecule sensitive to different components of external fields. 
%local operators, which only affect individual atoms, or to global operators, which affect jointly all of the atoms in the molecule. 
As an example, we can consider molecules with magnetic dipole moments versus molecules with only magnetic quadrupole moments. Magnetic dipoles are sensitive to uniform magnetic fields while magnetic quadrupoles are only sensitive to magnetic field gradients. In natural molecules, however, the order of the magnetic moment cannot easily be changed.

Here, we report an experiment in which we build an artificial molecule whose magnetic moment can be tuned via an applied external flux. The artificial molecule is composed of two fluxonium atoms strongly coupled via a shared inductance. At low applied external flux, the ground state $\g$ consists of persistent currents flowing in the same direction while the excited state $\e$ is an odd superposition of persistent currents flowing in opposite directions. The $\{\g,\e\}$ manifold is predominantly sensitive to common-mode flux noise that affects both atoms simultaneously. As the applied external flux is increased, $\g$ changes character and becomes the even superposition of persistent currents flowing in opposite directions while $\e$ essentially retains its symmetry. While this $\{\g,\e\}$ manifold has become insensitive to common-mode flux noise, it remains sensitive to differential-mode flux noise. We perform spectroscopy on this molecule and observe transitions between the ground state and excited states over multiple quanta of applied external flux. We investigate decoherence mechanisms in the device and conclude that the coherence of the $\g-\e$ transition is limited by local flux noise.

\section{Theoretical Model}
\begin{figure}[!t]
\includegraphics[width=86mm]{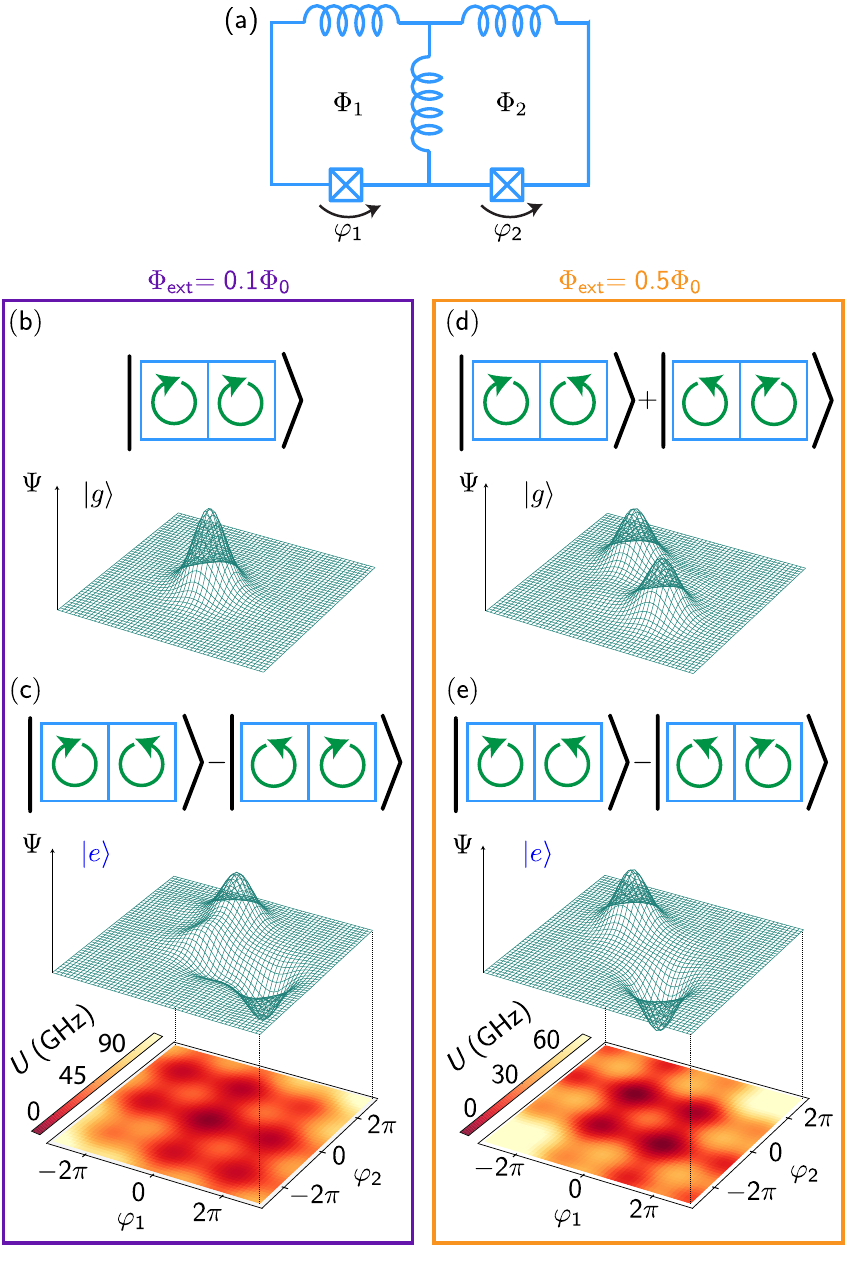}
\caption{(a) Electrical circuit diagram of the artificial molecule. (b-c) Magnetic dipole regime: Ground and excited-state wavefunctions ($\Psi$), and potential ($U$) at $\fqn=\overline{(\fqo+\fqt)}/2=0.1\Phi_0$. The direction of persistent current flow in the ground state is indicated with green arrows. The ground state is a product state of the two fluxonium atoms and is localized in the potential well near $\varphi_1,\varphi_2=0$. The excited state corresponds to persistent currents flowing in opposite directions and is delocalized in multiple potential wells. (d-e) Magnetic quadrupole regime: approximate kets for the ground and excited states at $\fqn=0.5\Phi_0$. Ground-state and excited-state wavefunctions ($\Psi$) are shown above the potential ($U$) of the artificial molecule. The ground and excited states are symmetric and antisymmetric superpositions of persistent currents flowing in opposite directions. The wavefunctions are localized in the two lowest potential wells, which are degenerate. Higher-energy excited states tend to be localized in the shallower potential wells. \label{fig:circuit}}
\end{figure}

Our system is composed of two fluxonium atoms coupled via a shared inductance as shown in Fig.~\ref{fig:circuit}(a). The artificial molecule obeys the Hamiltonian:
\begin{align}
%\begin{split}
H&=4 E_C(n_1^2+n_2^2)+\frac{1}{3}E_L(\varphi_1^2+\varphi_2^2+\varphi_1\varphi_2)\nonumber\\
& -E_J \cos\left(\varphi_1-\frac{2\pi}{\Phi_0}\Phi_1\right)-E_J \cos\left(\varphi_2-\frac{2\pi}{\Phi_0}\Phi_2\right),
%\end{split}
\end{align}
where $E_C$ is the charging energy of each junction, $n_1$ and $n_2$ are the number of Cooper pairs on each junction capacitor plate, $E_L$ is the inductive energy associated with each superinductance, $E_J$, $\varphi_1$, $\varphi_2$ are, respectively, the Josephson energies and the phase differences across each of the junctions, $\Phi_1$ and $\Phi_2$ are the fluxes in each loop, and $\Phi_0$ is the magnetic flux quantum. Here, we have assumed that the small junctions are identical. The $\varphi_1 \varphi_2$ term reflects the inductive coupling in the circuit and corresponds to its ``molecular'' aspect. This coupling term has a positive sign and favors opposite phase differences across the two Josephson junctions. The two fluxonium atoms are strongly coupled; the strength of the coupling term in the artificial molecule is equal to the inductive energy of each individual fluxonium atom ($E_L/h\sim1~$GHz). 

With the help of a gauge transformation (i.e., a shift of the phases $\varphi_i$ to $\varphi_i+(2\pi/\Phi_0)\Phi_i, i=1,2$), we can rewrite Eq.~1 as:
\begin{align}
%\begin{split}
H&=4 E_C(n_1^2+n_2^2)\nonumber\\
&+\frac{1}{4}E_L\left[(\varphi_1+\varphi_2+\varphi_\text{com})^2+\frac{1}{3}(\varphi_1-\varphi_2+\varphi_\text{diff})^2 \right]\nonumber\\
& -2 E_J \cos\left(\frac{\varphi_1+\varphi_2}{2}\right)\cos\left(\frac{\varphi_1-\varphi_2}{2}\right), 
%\end{split}
\end{align}
where we have introduced the common and differential applied reduced fluxes, $\varphi_\text{com}=\frac{2 \pi}{\Phi_0}(\Phi_1+\Phi_2)$ and $\varphi_\text{diff}=\frac{2\pi}{\Phi_0}(\Phi_1-\Phi_2)$. This form of the Hamiltonian illustrates the sensitivity of the molecule to both modes of magnetic field fluctuations.

The artificial molecule is operated in two regimes by changing an applied external magnetic flux, $\fqn$, which sets the temporal average of the common-mode flux $\overline{(\Phi_1+\Phi_2)}/2=\fqn$, and, to a much weaker extent, the temporal average of the differential-mode flux $\overline{(\Phi_1-\Phi_2)}=\alpha \fqn$, where $\alpha \ll 1$. In the first regime, $0\leq \fqn \lesssim 0.3 \Phi_0$, the molecule essentially behaves like a magnetic dipole. The phase difference across both of the junctions in the molecule is the same. As shown in Fig.~1(b,c), the potential has one deep well centered near $\varphi_1=0$ and $\varphi_2=0$, and the ground state is localized in this well. This ground state is simply the product of the single fluxonium-atom ground states and corresponds to currents flowing in the same direction in both fluxonium atoms. The direction of current flow is determined by the direction of the applied external flux. It helps to think of the persistent current chirality as the spin of the fluxonium atom. The excited state, then, corresponds to the singlet spin state while the ground state corresponds to the $m=1$ triplet spin state. To transition from $\g$ to $\e$, the persistent current in one of the fluxonium atoms needs to change direction, which is analogous to flipping a spin. The transition frequency here is then determined by the applied external flux and is, hence, sensitive to noise in the common-mode flux.

\begin{table*}
\begin{ruledtabular}
\begin{tabular}{ccccccccc}
 &  & & &&\multicolumn{4}{c}{$\fq{0.5}$}\\\cline{6-9}
Device & $E_J \text{(GHz)}$ &$E_C \text{(GHz)}$& $E_L \text{(GHz)}$ &$\alpha$& $f_{ge} \text{(MHz)}$ &$T_1 (\mu \text{s})$ &$T_{2R} (\mu \text{s})$ &$T_{2E} (\mu \text{s})$ \\
\hline
A & $9.4$ & $3.4$ & $1.2$ & $0.006$ & $105$ & $60$ & $0.45$ & $1.6$ \\
%\hline
B & $9.5$ & $3.4$ & $1.1$ & $0.007$ & $110$ & $50$ & $1.25$ & $5.4$ \\
%\hline
C & $9.8$ & $3.3$ & $1.2$ & $0.03$ & $197$ & $300$ & $0.31$ & $1.2$ \\
%\hline
\end{tabular}
\end{ruledtabular}
\caption{Parameters of three different artificial molecule devices. Here, $E_J$ is the Josephson energy of each fluxonium atom, $E_C$ is the charging energy of each fluxonium atom, $E_L$ is the inductive energy associated with each superinductance, $\alpha$ is the asymmetry between the fluxes of each fluxonium atom, $f_{ge}$ is the molecule transition frequency from $\g$ to $\e$ at $\fq{0.5}$, $T_1$ is the relaxation time from $|e\rangle~\mathrm{to}~|g\rangle,~T_{2R}$ is the coherence time measured by a Ramsey experiment, and $T_{2E}$ is the coherence time measured by a spin-echo experiment. All time-domain measurements are done at $\fq{0.5}$. \label{tab:params}}
\end{table*}

As the external flux is increased, the two fluxonium atoms start to behave like a molecule with no dipole moment but with a quadrupole moment. In this regime, the potential landscape of the molecule has two degenerate potential wells as shown in Fig.~1(e). The degeneracy of the wells comes from the symmetry between $\varphi_1$ and $\varphi_2$ in the Hamiltonian (Eq.~1). These two wells correspond to the two possible configurations of persistent currents flowing in opposite directions in each loop. The energy splitting between $\g$ and $\e$ is determined by the tunneling between the two lowest degenerate potential wells. 

The ground-excited state manifold of the molecule near $\fq{0.5}$ is formed from superpositions of the currents flowing in opposite directions in the two fluxonium atoms as shown in Fig.~1(d-e). The $\g$ state thus corresponds to the $m=0$ triplet spin state and the $\e$ state corresponds to the singlet spin state. This ground-excited state manifold is gradiometric; the transition is insensitive to fluctuations in external flux that are uniform across both loops. We note that in the limit of perfect symmetry between the two fluxonium atoms, this ground-excited state manifold is also insensitive to first order to fluctuations in external flux that occur only in individual loops.

\begin{figure}
\includegraphics[width=86mm]{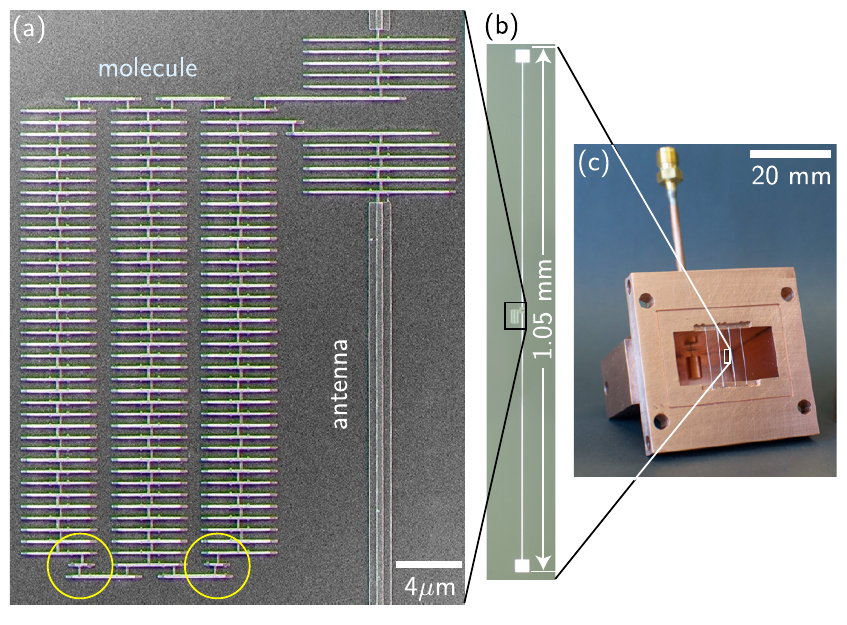}
\caption{(a) Scanning electron micrograph of device A. Small junctions are indicated in the yellow circles. The artificial molecule is coupled via Josephson junctions to the readout antenna whose optical image is shown in (b). (c) The molecule and antenna are fabricated on a sapphire chip, which is then placed inside of a copper waveguide. \label{fig:device}}
\end{figure}

As a consequence of this insensitivity, $f_{ge}$ is nearly constant over a large range in $\fqn$, which should protect this transition from common-mode flux noise. Another way to see this is to consider that both junctions must undergo a quantum phase slip in order to tunnel between the two lowest potential wells. The persistent currents in both fluxonium atoms must flip direction together. A double phase slip is a second-order process where the molecule has to make a virtual transition through a higher-energy state with the currents circulating in the same direction in the two fluxonium loops. The energy splitting is then on the order of $E_S^2/\Delta$, where $E_S\sim(E^3_J E_C)^{1/4}e^{-\sqrt{8E_J/E_C}}$ is the energy of a single phase slip in an individual fluxonium atom \cite{Matveev2002,Meier2015} and $\Delta\sim \frac{2}{3} \pi^2 E_L$ is the energy difference between the states with counter-circulating currents in the two loops and the states with currents flowing in the same direction in the two loops.  We note here that the decoherence rate associated with an unwanted phase slip will be suppressed by a factor of $E_S/\Delta$ when compared with the single fluxonium atom.

\section{Experimental Realization}
The artificial molecule device (device A) is shown in Fig.~\ref{fig:device}(a). Each fluxonium atom consists of a small Josephson junction, which provides nonlinearity, in series with an array of 40 larger junctions. The array of 40 junctions has an inductance of 140~nH and serves as the superinductance for each atom \cite{superinductanceNote}. The fluxonium atoms are connected in parallel with an additional array of 40 junctions, which provide inductive coupling between the atoms. The artificial molecule is inductively coupled to a readout antenna via shared Josephson junctions, which have an inductance of 3.0~nH.

The inductively-loaded readout antenna is an LC oscillator where the inductance is provided by 14 Josephson junctions and the capacitance is provided by the long metal electrodes as shown in Fig.~\ref{fig:device}(b). The antenna has a resonant frequency of $f_a=7.875~$GHz and a FWHM linewidth of $\kappa/2\pi=6~$MHz. The junctions were fabricated with Al/AlO$_x$/Al using the bridge-free double-angle evaporation technique \cite{Lecocq2011}. Both the molecule and the antenna were fabricated on a sapphire chip. This chip was then placed inside of an impedance-matched copper waveguide (Fig.~\ref{fig:device}(c)), which couples propagating microwaves to the molecule-antenna system. The waveguide was thermally anchored to the mixing chamber stage of a dilution refrigerator with a base temperature of $\sim16~$mK. The waveguide was magnetically shielded by an aluminum shield coated with infrared-absorbing material, which was itself housed inside of a cryogenic $\mu$-metal shield \cite{KurtisThesis}.
\begin{figure*}[!t]
\includegraphics[width=172mm]{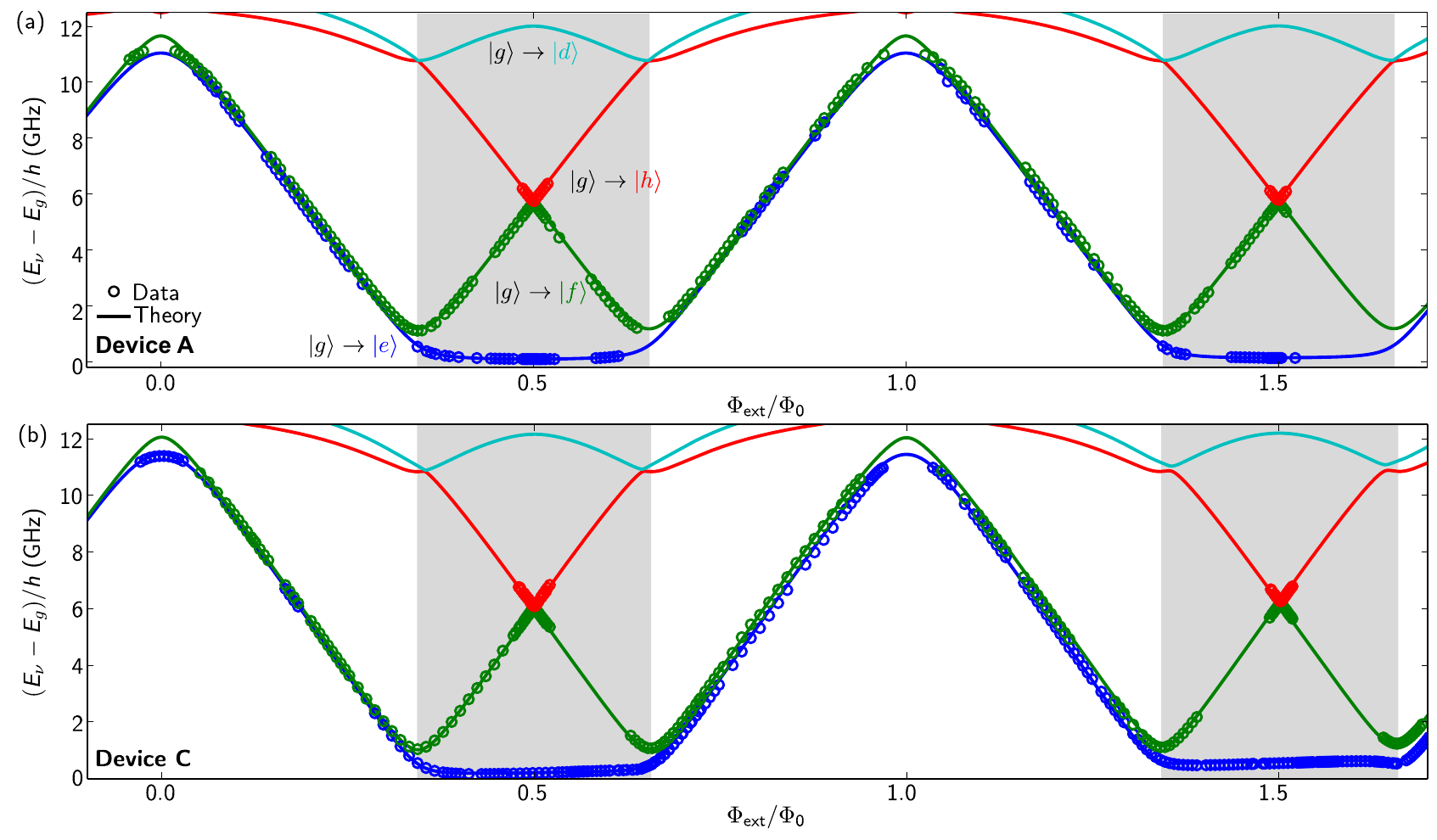}
\caption{(a-b) Transition frequency from $\g$ to excited states $\e,|f\rangle,|h\rangle,|d\rangle$ in device A and device C. Measured data is indicated with open circles. The solid lines were fit to the data using the Hamiltonian in Eq.~1 and the definition of $\fqn$ (see text). Grey shaded areas indicate where the $\g-\e$ transition is only quadrupolar. Device C has a higher asymmetry between the two loops, which leads to a larger difference between the $\g$ to $\e$ transition frequency at $\fq{0.5}$ and $\g$ to $\e$ transition frequency at $\fq{1.5}$ when compared with device A. \label{fig:spec}}
\end{figure*}

\section{Molecular Spectroscopy}
We used standard dispersive readout \cite{Blais2004} to measure the molecule with the readout antenna. We performed a two-tone spectroscopy experiment at different flux points to obtain the data shown in open circles in Fig.~\ref{fig:spec}. We first applied a fixed $\fqn$ through the device and then shone a microwave tone to excite the molecule at its resonant frequency, which resulted in a change in the resonant frequency of the readout antenna. The sample was biased via a large magnetic field coil that encircles the entire device. This experiment was then repeated at multiple flux points between $\Phi_\mathrm{ext}\sim -0.1 \Phi_0$ and $\Phi_\mathrm{ext} \sim 1.5 \Phi_0$. We observed transitions between $\g$ and the first three excited states. 

At the critical point (indicated by the start of the grey shaded area in Fig.~\ref{fig:spec}) at $\Phi_\mathrm{ext} \approx 0.3 \Phi_0$, we observed the transition frequency from $\g$ to $|f\rangle$ start to increase as a function of flux. For $\fqn>0.3\Phi_0$, the $\g-\e$ transition frequency, $f_{ge}$, becomes very small. At its lowest point, $f_{ge}$ reaches 105~MHz at $\fq{0.5}$. In contrast with other flux-based qubits such as the flux qubit \cite{VanderWalCH1999} and the fluxonium \cite{Manucharyan2009}, $f_{ge}$ is also remarkably flat as a function of $\fqn$ near $\fq{0.5}$, as shown in Fig.~\ref{fig:spec} \cite{Gladchenko2009}.

At $\fq{0.3}$, the potential landscape of the system transitions from having a single lowest potential well to having two degenerate potential wells. The former ground state of the molecule -- currents circulating in the same direction in both loops -- becomes the $|f\rangle$ state. For $\fqn>0.3\Phi_0$, the $\g$ and $\e$ states are superpositions of currents flowing in opposite directions in the two loops as shown in Fig~\ref{fig:circuit}(d,e). The energy splitting between $\g$ and $\e$ in this regime is $\sim E_S^2/\Delta$, which results in the very small $f_{ge}$. The flatness of this transition as a function of flux results from the necessity of using a higher-energy state to transition between the two lowest potential wells.

We also note that the artificial molecule has different $\g-\e$ transition frequencies at external flux points separated by a flux quantum, which is to be expected in a device with unequal fluxonium loop areas. The $\g-\e$ transition frequency at $\fq{1.5}$ is 40~MHz higher than $f_{ge}$ at $\fq{0.5}$ in device A. 

Differences in the flux through the two fluxonium loops result in $\overline{\fqo-\fqt}=\alpha\fqn\neq 0$ in the Hamiltonian (Eq.~(1)). The two fluxonium atoms then have different potential energies at each flux point. As the applied external flux is increased, this difference in potential landscapes compounds. The potential wells corresponding to currents flowing in opposite directions in the two loops are then no longer close to degenerate, resulting in the observed larger energy splitting at $\fq{1.5}$.

We compare the measured resonant frequencies (open circles in Fig.~\ref{fig:spec}) with the theoretical prediction for these transition frequencies (solid lines in Fig.~\ref{fig:spec}) obtained from numerical diagonalization of the Hamiltonian (Eq.~(1)) \cite{Smith2016}. The only asymmetry we have considered here is the nonzero $\alpha$ parameter. To fit the data, we constrain the product $E_J E_C$ based on fabrication parameters and effectively fit the full spectrum using only three fit parameters, $\{\alpha,E_J/E_C,E_L\}$. We find excellent agreement between the model and the data over three decades in transition frequencies (from 11.2~GHz at $\fqn=0$ to 105~MHz at $\fq{0.5}$). This validates the approach of planning the behavior of a complex circuit, such as the one shown in Fig.~\ref{fig:device}, from a reduced set of engineerable parameters.

We further tested the validity of our approach by measuring multiple devices. The fit parameters for the measured devices are shown in Tab.~\ref{tab:params}. The spectroscopy of device C, which intentionally had the largest asymmetry between the two fluxonium loops, is shown in Fig.~\ref{fig:spec}(b). For this device, the fit parameter $\alpha$ is consistent with the asymmetry inferred from optical images. We note again that the agreement between the theoretical fit and the measured data here is excellent. The effects of asymmetry are much more pronounced in this device. The lowest $f_{ge}=197~\text{MHz}$ no longer occurs at $\fq{0.5}$ but rather at $\fq{0.43}$. In addition, $f_{ge}$ at $\fq{1.5}$ is now 354~MHz greater than $f_{ge}$ at $\fq{0.5}$. 

\section{Time Domain Measurements}
In order to demonstrate the insensitivity of the molecule to certain decoherence mechanisms, we measured the coherence times of the $\g-\e$ transition for the three devices (A, B, and C) near $\fq{0.5}$. We performed standard time-domain measurements of the relaxation time ($T_1$), Ramsey dephasing time ($T_{2\mathrm{R}}$), and spin-echo dephasing time ($T_{2\mathrm{E}}$). The coherence times at $\fq{0.5}$ are shown in Tab.~\ref{tab:params}. Relaxation times for the measured devices are between $50-300~\mu$s. Ramsey coherence times for the measured devices are between $0.4-1.25~\mu$s. Spin echo experiments, where a $\pi-$pulse is inserted into the standard Ramsey sequence, increased the coherence times by a factor of 4, indicating the presence of a low-frequency decoherence mechanism.

\begin{figure}[t]
\includegraphics[width=86mm]{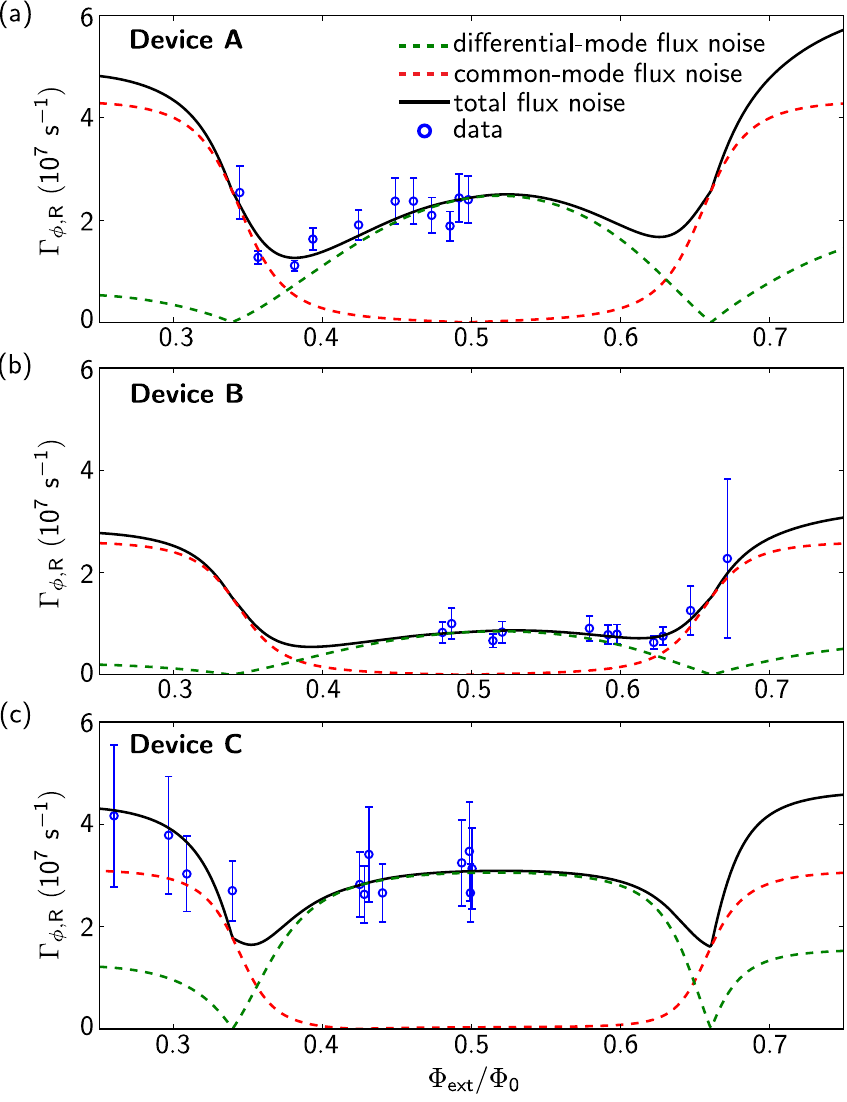}
\caption{(a-c) Ramsey dephasing rates as a function of flux in device A, B, and C. The data is shown in open blue circles. Fit curves are calculated Ramsey dephasing rates resulting from power spectral density values of differential-mode flux noise (green dashed), common-mode flux noise (red dashed), and their sum (black solid). We note that slight asymmetries in the two fluxonium loops cause the curves to be asymmetric around $\fq{0.5}$. This asymmetry also affects the amplitudes necessary to drive transitions and the visibility of the transitions between $\g$ and $\e$, resulting in the asymmetry in data about $\fq{0.5}$ in these plots. Ramsey dephasing rate measurements could not be taken for the full range in $\fqn$ for every sample due to changing coupling between the molecule and readout antenna. \label{fig:ramsey}}
\end{figure}

In order to understand the mechanisms for decoherence in this molecule, which has surprisingly low $T_2$'s given $T_1$, we measured the Ramsey dephasing rate, $\Gamma_{\phi,R}$ as a function of $\fqn$. The dependence of the $f_{ge}$ transition on $\fqn$ allows us to isolate the contributions of common-mode and differential-mode flux noise to dephasing of the molecule as shown in Fig.~\ref{fig:ramsey} (see Suppl.~material for details). Near $\fq{0.5}$, the coherence of the $\g-\e$ transition is limited by differential-mode flux noise while at $\Phi_\text{ext}\lesssim 0.3$ it is predominantly limited by common-mode flux noise.

The spectral density of noise resulting from an arbitrary source $\lambda$ is,
\begin{equation}
\begin{split}
S_{\delta\lambda}(\omega)=\frac{1}{2\pi}\int_{-\infty}^{\infty}d\tau \langle \delta\lambda(t)\delta\lambda(t+\tau)\rangle e^{-i\omega\tau},
\end{split}
\end{equation}
where $\omega$ is the frequency at which the spectral density is taken. For common-mode flux noise, $\lambda=\frac{\Phi_0}{2\pi\sqrt{2}}\varphi_\mathrm{com}$ and for differential-mode flux noise, $\lambda=\frac{\Phi_0}{2\pi\sqrt{2}}\varphi_\mathrm{diff}$.

Flux noise is typically assumed to have a $1/f$ spectrum, i.e. $S=A^2/|\omega|$, where $A$ is the flux noise amplitude. The common-mode and differential-mode flux noise amplitudes are given in Tab.~\ref{tab:fluxnoise}. We find upper bounds for flux noise amplitudes between $4-11~\mu\Phi_0$. The flux noise amplitudes that we measure are somewhat larger but of the same order of magnitude as previous measurements of flux noise in flux qubits \cite{Yoshihara2006,Bylander2011} and coupled flux qubits \cite{Yoshihara2010}. 

\begin{table}[!tb]
\begin{ruledtabular}
\begin{tabular}{c c c c c}
 &  & & \multicolumn{2}{c}{$A_\text{diff}/A_\text{com}$}\\\cline{4-5}
Device & $A_\text{com}~(\mu\Phi_0)$ & $A_\text{diff}~(\mu\Phi_0)$ & Exp. Global & Exp. Local \\
\hline
A & $6 \pm 1$ & $10 \pm 2$ & 0 & $>1$ \\
%\line
B & $4 \pm 1$ & $4 \pm 1$ & 0 & $>1$  \\
%\hline
C & $8 \pm 1$ & $11 \pm 2$ & 0 & $>1$\\

\end{tabular}
\end{ruledtabular}
\caption{Inferred flux noise amplitudes from $T_{2R}$ measurements together with the expected ratio of differential-mode to common-mode flux noise amplitudes for each type of flux noise. \label{tab:fluxnoise}}
\end{table}

Global flux noise, from a fluctuating applied magnetic field for example, would induce exclusively common-mode flux noise in the sample ($A_\text{diff}/A_\text{com}=0$). On the other hand, local flux noise, such as that caused by spins fluctuating at the surface of the superconductor \cite{Koch2007} or in defects at the metal/insulator interface \cite{Faoro2008}, induces both common-mode and differential-mode flux noise in the sample. The exact branching ratio depends on the details of the model but, in general, the differential-mode contribution would be greater than the common-mode contribution. Hence, for local flux noise, we would expect $A_\text{diff}/A_\text{com}>1$. We observe the latter behavior in all three devices. Near $\fq{0.5}$, the sample is insensitive to common-mode flux noise and is limited by differential-mode flux noise, which can only result from local sources.

We note that the residual asymmetry between the two loops in the molecule results in sensitivity to differential-mode flux noise (see Suppl.~Material). A locally-applied external flux could be used to compensate this asymmetry, which may result in longer coherence times. In addition, surface passivation has recently been found to decrease flux noise in superconducting quantum interference devices \cite{Kumar2016} and may also increase coherence times.

\section{Conclusion}
In conclusion, we have realized an artificial molecule based on the fluxonium atom, which possesses a tunable magnetic moment. A very basic circuit model accounts for all of the details observed in the spectroscopy of the molecule. The $\g-\e$ transition of this molecule near $\fq{0.5}$ is robust to excursions in common-mode flux, whereas for $\Phi_\mathrm{ext}\lesssim0.3\Phi_0$ the transition is sensitive to excursions in both common-mode and differential-mode flux. 

Surprisingly, the coherence of the ground-excited state manifold near $\fq{0.5}$ is not improved by this insensitivity to common-mode flux noise. Detailed measurements of $\Gamma_{\phi,R}$ as a function of $\fqn$ pinpoint local flux noise as the main limitation for the coherence of these devices. Our work thus constitutes new evidence for the hypothesized local flux noise in superconducting circuits. A possible solution to this local source of decoherence are alternative fluxonium-based circuits, which would be practical realizations of the proposed ideas of Ioffe et al. \cite{Ioffe2002} and Brooks et al. \cite{Brooks2013}. 
%Further work is needed to understand the origin of local flux noise in superconducting circuits.

\section{Acknowledgements}
We acknowledge fruitful discussions with Rob Schoelkopf, Ioan Pop, Shyam Shankar, and Chen Wang. Facilities use was supported by the Yale SEAS cleanroom, YINQE and NSF MRSEC DMR-1119826. This research was supported by the Army Research Office under Grant No. W911NF-14-1-0011 and by Office for Naval Research under Grant No. N00014-16-1-2270.. We  acknowledge additional support from NSF DMR-1609326 (S.M.G), NSF DMR-1301798 (S.M.G), and DMR-1603243 (L.G.). 

\section*{Appendix I: Decoherence Mechanisms}
\subsection{Flux Noise}
In the main text, we argued that local flux noise is the primary source of decoherence in our device. Here, we provide theoretical details for the curves plotted in Fig.~4. We first fit the spectrum of the molecule to determine $f_{ge}$ as a function of $\fqn$ by numerically diagonalizing the molecule Hamiltonian including corrections for asymmetry:

\begin{equation}
\begin{split}
H&=4 E_C(n_1^2+n_2^2)+\frac{1}{3}E_L(\varphi_1^2+\varphi_2^2+\varphi_1\varphi_2)\\
& -E_J \cos \left (\varphi_1-\left (1+\frac{\alpha}{2} \right )2 \pi\Phi_\mathrm{ext}/\Phi_0 \right)\\
& -E_J \cos \left (\varphi_2- \left (1-\frac{\alpha}{2} \right )2\pi\Phi_\mathrm{ext}/\Phi_0 \right ),
\end{split}
\end{equation}
where $E_C$ is the charging energy of each junction, $n_1$ and $n_2$ are the number of Cooper pairs on each junction capacitor plate, $E_L$ is the inductive energy of each superinductance, $\varphi_1$ and $\varphi_2$ are the phase differences across each junction, $E_J$ is the Josephson energy of each junction, $\alpha$ is the difference in the fluxes of the two fluxonium qubits, and $\Phi_\mathrm{ext}$ is the applied external flux in each loop.

To calculate the common-mode flux noise limits, we then take a derivative of $f_{ge}$ as a function of $\fqn$ using finite difference methods. Following the analysis of Ithier et al. \cite{Ithier2005}, we fit the Ramsey coherence data using the following equation:
 \begin{equation}
\Gamma_{\phi,R}=2\pi\sqrt{A_\text{com}\eta_R}\left |\frac{\partial f_{ge}}{\partial \Phi_\mathrm{ext}} \right |,
\end{equation}
where $A_\text{com}$ is the common-mode flux noise amplitude, $\eta_R=\ln[\Gamma_{\phi,R}/(2 \pi f_\text{ir})]$, which depends on the low-frequency cutoff for the flux noise, $f_\text{ir}$. We take $f_\text{ir}=1$ Hz, which corresponds to the inverse of our averaging time. This cutoff is consistent with the measured ratio $\Gamma_{\phi,R}/\Gamma_{\phi,E}=\sqrt{\ln[\Gamma_{\phi,R}/(2 \pi f_\text{ir})]/\ln 2} \sim 4$. From this fit we obtain the common-mode flux noise amplitudes given in Tab.~II.

To calculate the differential-mode flux noise limits, we take a derivative of $f_{ge}$ as a function of of $\alpha/2$ again using finite difference methods. We fit the $T_{2R}$ data using the equation: 
\begin{equation}
\Gamma_{\phi,R}=2\pi\sqrt{A_\text{diff}\eta_R}\frac{1}{\fqn}\left |\frac{\partial f_{ge}}{\partial (\alpha/2)} \right |,
\end{equation}
where $A_\text{diff}$ is the differential-mode flux noise amplitude and $\eta_R$ is the same constant as above. From this fit we obtain the differential-mode flux noise amplitudes given in Tab.~II.

\begin{figure}[t]
\includegraphics[width=86mm]{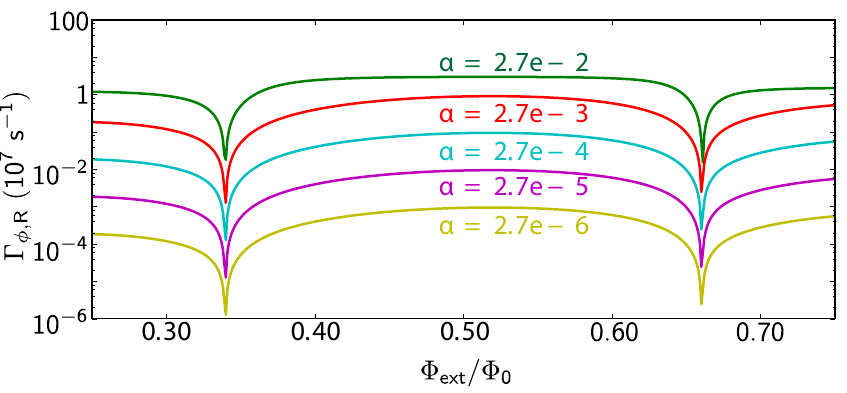}
\caption{Calculated Ramsey dephasing rates as a function of $\fqn$ for various asymmetries between the fluxes of the two fluxonium atoms using the parameters from device C. \label{fig:asymm}}
\end{figure}

We remark that a stronger residual asymmetry between the two fluxes results in lower coherence times of the molecule around $\fq{0.5}$. Figure~\ref{fig:asymm} shows the calculated $\Gamma_{\phi,R}$ limits from differential-mode flux noise for the parameters of device B with decreasing asymmetry between the the fluxes of the two fluxonium atoms.

\subsection{Critical Current Noise}
\begin{figure}[t]
\includegraphics[width=86mm]{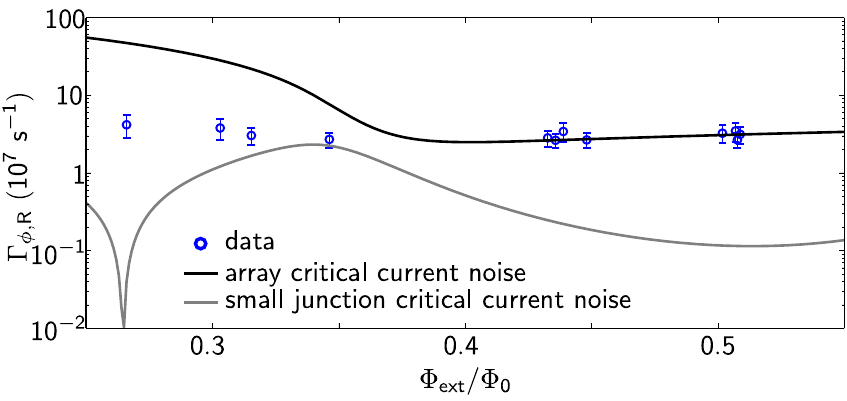}
\caption{Ramsey dephasing rates measured for device C shown in blue circles. The grey solid line indicates the limit on $\Gamma_{\phi,R}$ due to critical current noise from the small junctions using the parameters of device C. The black solid line indicates the limit on $\Gamma_{\phi,R}$ due to critical current noise from junctions in the array using the parameters of device C. Both limits correspond to a critical current noise amplitude of $A_{I_C}=10^{-3} I_C$, where $E_{J}=\frac{\hbar}{2e}I_C$ is the energy of each junction. \label{fig:el_dependence}}
\end{figure}

Critical current noise corresponds to noisy excursions in the Josephson energy of each of the junctions in the molecule. Near $\fq{0.5}$, the energy splitting between $\g$ and $\e$ is on the order of $E_S^2/\Delta$, where $E_S\sim(E_J^3 E_C)^{1/4} e^{-\sqrt{8 E_J/E_C}}$ and $\Delta \sim \frac{2}{3} \pi^2 E_L$. The contribution to dephasing from the small junctions is then suppressed by $E_S/\Delta\sim0.1$ when compared with the contribution to dephasing from critical current noise in the small junction for a single fluxonium qubit \cite{Manucharyan2009}. The solid grey line in Fig.~\ref{fig:el_dependence} shows the calculated $\Gamma_{\phi,R}$ limit as a function of $\fqn$ due to critical current noise in the small junctions in device C. We note that the curve corresponds to a noise amplitude of $A_{I_C}=10^{-3} I_C$, which is more than three orders of magnitude larger than measured in previous experiments \cite{Harlingen2004} and that the measured data does not obey the shape of this curve. This suggests that the coherence of the $\g-\e$ transition is not limited by critical current noise in the small junctions. 

Critical current noise will also induce fluctuations in the inductance of the molecule because the superinductors are composed of large Josephson junctions, however, the central limit theorem shows that the relative amplitude of the noise is suppressed by $\sqrt{N_A}$, where $N_A$ is the number of junctions in the superinductor. Figure \ref{fig:el_dependence} also shows the calculated $\Gamma_{\phi,R}$ limit as a function of $\fqn$ due to critical current noise in the array junctions in device C with $A_{I_C}=10^{-3} I_C$. We note again the extremely large value of $A_{I_C}$ necessary to fit the data. We therefore conclude that the molecule is not limited by critical current noise in the array junctions or the small junction. 

\begin{figure}[b]
\includegraphics[width=86mm]{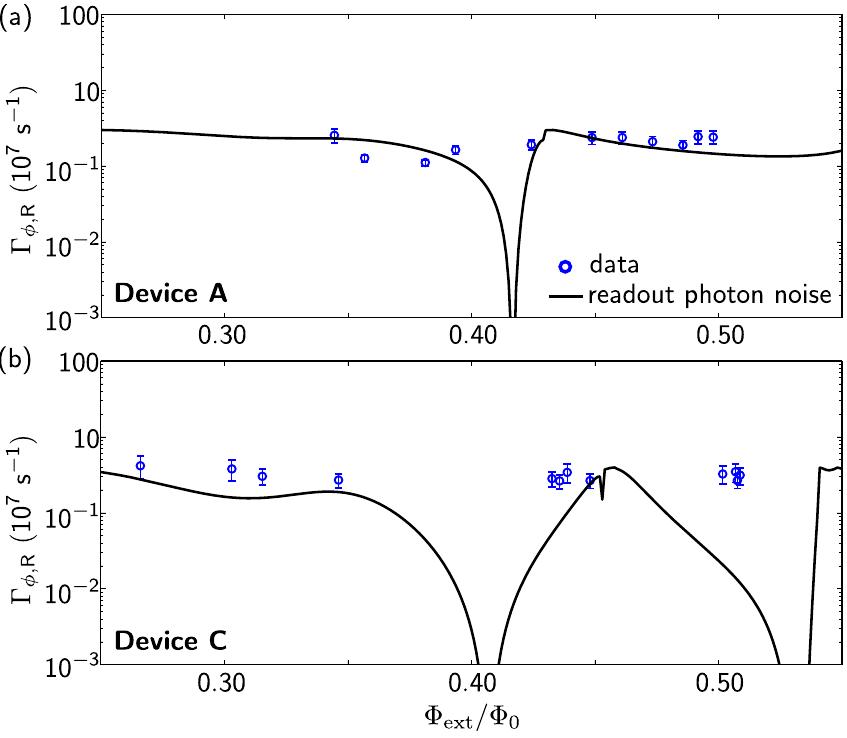}
\caption{Ramsey dephasing rates measured in (a) device A and (b) device C shown in blue circles. Readout photon noise limits on $\Gamma_{\phi,R}$ with $\bar{n}=0.5$ are shown as black solid lines. \label{fig:cavnoise}}
\end{figure}

\begin{figure*}
\includegraphics[width=172mm]{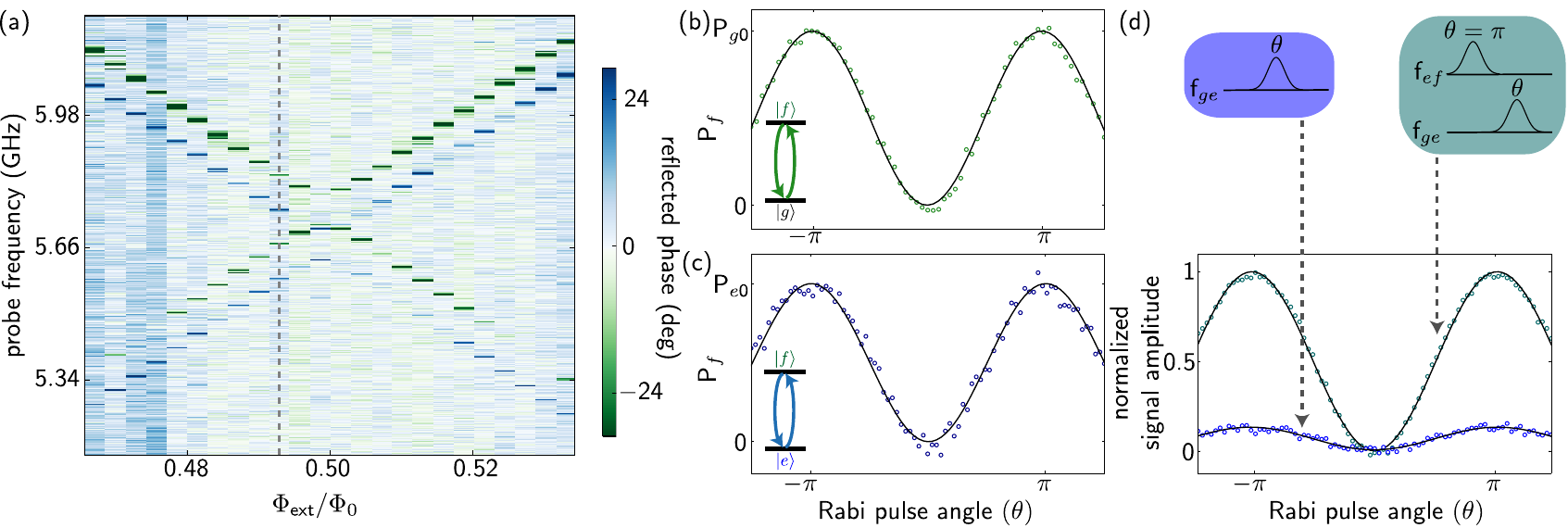}
\caption{(a) Spectroscopy of the $\{\g,\e\}$ manifold transitions to the higher-energy excited state manifold in device A. The transitions between $\g$ and $|f\rangle,|h\rangle$ cause an increase in the measured reflected phase of the readout antenna. The low frequency of the $\g-\e$ transition results in a large equilibrium thermal population in $\e$. This population causes transitions between $\e$ and $|f\rangle,|h\rangle$ to also be visible in spectroscopy; these transitions result in a decrease in the phase of the readout antenna. The grey dashed line indicates the location where we perform the Rabi experiments. (b) Rabi oscillations between $\g$ and $|f\rangle$. The applied drive transfers the population of $\g$ ($P_g$) to $|f\rangle$. c) Rabi oscillations between $\e$ and $|f\rangle$. The applied drive transfers the population of $\e$ ($P_e$) to $|f\rangle$. (d) The populations of $\g,\e$, and $|f\rangle$ can be controlled. Direct driving of the $\g$ to $\e$ state with no preparation pulse results in a small Rabi contrast due to the nearly equal populations in $\g$ and $\e$. The contrast can be increased if a preparation pulse shifting the population in $\e$ to the population in $|f\rangle$ is first applied. The signal amplitudes are normalized to the maximum amplitude obtained from the Rabi experiment with a preparation pulse. \label{fig:state}}
\end{figure*}

\subsection{Readout Antenna Photon Noise}
The presence of residual thermal photons in the readout antenna can lead to decoherence of the $\g-\e$ transition. The readout antenna is coupled to the molecule via a small shared inductance of $3.0~$nH. This coupling results in a dispersive shift of the readout antenna frequency depending on the state of the molecule and, conversely, shifts in the transition frequencies of the molecule depending on the instantaneous photon number ($n$) in the antenna. Fluctuations in $n$ will change the transition frequencies of the molecule, which lead to dephasing of the molecule's transitions. We can incorporate the readout antenna and its coupling into the Hamiltonian of Eq. 1 and find the $\Gamma_{\phi,R}$ limit for a given photon number via calculated energy levels. Assuming $\bar{n}\ll1$ and a thermal distribution for $n$,
 \begin{equation}
\Gamma_{\phi,R}=\bar{n} \kappa \chi^2 \frac{1}{\kappa^2+\chi^2},
\end{equation}
where $2\pi/\kappa$ is the lifetime of the readout antenna and $\chi=2\pi(f_{1,ge}-f_{0,ge})$ is the shift of the $\g-\e$ transition frequency when the readout antenna is populated with 1 photon \cite{Gambetta2006,Clerk2007,Yan2015}. This curve is plotted for device A and device C in Fig.~\ref{fig:cavnoise} with $\bar{n}=0.5$. This $\bar{n}$ corresponds to a $\g-\e$ temperature of 300~mK and is inconsistent with previous measurements of the fluxonium qubit temperature carried out in the same measurement setup \cite{unpubAkou}. The shape of the $\Gamma_{\phi,R}$ as a function of $\fqn$ when limited by photon noise is also inconsistent with the behavior of $\Gamma_{\phi,R}$ as a function of $\fqn$ so we conclude that the molecule is not limited by readout antenna photon noise.

\section*{Appendix II: Relaxation Mechanisms}
We note that the $T_1$'s for the devices measured are between $70-300~\mu$s. Previous measurements of the fluxonium qubit coupled to a cavity found $T_1$'s of up to 1~ms \cite{Pop2014}. Calculations of the Purcell loss through the antenna have predicted lifetimes in excess of $1~\text{ms}$ so we do not think that direct coupling to the waveguide-termination impedance is responsible for this difference in lifetimes. Possible culprits include increased dielectric losses in these samples, excess quasiparticle dissipation, and the increased exposure to noise at frequencies greater than $8~\text{GHz}$ in these experiments that use an open waveguide rather than a closed cavity.

\section*{Appendix III: Population Control via Higher Excited States}

Transitions to high-level states have been used  successfully in natural atoms \cite{Ludlow2006} and ions \cite{Langer2005} to readout and manipulate low-frequency manifolds. In addition, in future implementations of qubits protected from all local decoherence mechanisms, the qubit transition cannot be addressed directly and will require the use of higher-energy states for readout and manipulation. The unique level structure of our molecule allows us to use higher excited states for manipulation of the nearly-degenerate $\{\g,\e\}$ manifold near $\fq{0.5}$. Figure~\ref{fig:state}(a) shows spectroscopy of transitions from the $\{\g,\e\}$ manifold to higher excited states in device A. The small transition frequency, $f_{ge}$, of the molecule resulted in a thermal population of $\approx 45\%$ at the base temperature of the dilution refrigerator. Hence, when we applied microwave drives, we excited population from both $\g$ and $\e$ into the higher-energy states. We observed changes in the antenna response for both $\g$ and $\e$ transitions to the second ($| f\rangle$) and third ($| h \rangle$) excited state. 

We also demonstrate that we could coherently manipulate the population between $\g$ and $| f\rangle$ as shown in Fig.~\ref{fig:state}(b). Here, we applied a microwave tone with a pulse width of 20~ns at $f_{gf}$ and observed Rabi oscillations between $\g$ and $| f\rangle$. We could similarly manipulate the population between $\e$ and $| f\rangle$ as shown in Fig.~\ref{fig:state}(c). We then applied a microwave tone with a pulse width of 20~ns at $f_{ef}$ to observe Rabi oscillations between $\e$ and $| f\rangle$. While we could coherently transfer population between $\g-|f\rangle$ and $\e-|f\rangle$, stimulated Raman adiabatic passage \cite{STIRAPreview} in these devices would be less efficient than direct population transfer through $|f\rangle$ because the coherence times of the $\g-|f\rangle$ and $\e-|f\rangle$ transitions are not significantly shorter than the coherence time of the $\{\g,\e\}$ manifold. 

By transferring population from the $\{\g,\e\}$ manifold to $| f\rangle$, we can significantly increase the readout contrast of the $\g-\e$ transition. The low frequency of the $\g-\e$ transition results in a large equilibrium thermal population in $\e$. Therefore, when a $\pi-$pulse is applied to drive the $\g-\e$ transition, the population inversion only results in a small change in the signal amplitude as shown in Fig.~\ref{fig:state}(c). The readout contrast increases if we first apply a $\pi-$pulse on the $\e-| f\rangle$ transition to transfer the population from $\e$ to $| f\rangle$ and then apply a $\pi-$pulse on the $\g-\e$ transition as shown in Fig.~\ref{fig:state}(d).  

While this method was not used for the data acquisition of Figs.~3 and 4, the ability to readout and manipulate the $\g-\e$ manifold via higher excited states would be useful in future designs where the $\g-\e$ transition no longer has to be directly accessed. This would allow for further protection of the molecule from decoherence mechanisms such as the presence of residual thermal photons in the readout antenna \cite{Gambetta2006}.

\bibliographystyle{apsrevx}
\bibliography{dfluxonium}

%merlin.mbs apsrev4-1.bst 2010-07-25 4.21a (PWD, AO, DPC) hacked
%Control: key (0)
%Control: author (72) initials jnrlst
%Control: editor formatted (1) identically to author
%Control: production of article title (-1) disabled
%Control: page (0) single
%Control: year (1) truncated
%Control: production of eprint (0) enabled
\begin{thebibliography}{33}%
\makeatletter
\providecommand \@ifxundefined [1]{%
 \@ifx{#1\undefined}
}%
\providecommand \@ifnum [1]{%
 \ifnum #1\expandafter \@firstoftwo
 \else \expandafter \@secondoftwo
 \fi
}%
\providecommand \@ifx [1]{%
 \ifx #1\expandafter \@firstoftwo
 \else \expandafter \@secondoftwo
 \fi
}%
\providecommand \natexlab [1]{#1}%
\providecommand \enquote  [1]{``#1''}%
\providecommand \bibnamefont  [1]{#1}%
\providecommand \bibfnamefont [1]{#1}%
\providecommand \citenamefont [1]{#1}%
\providecommand \href@noop [0]{\@secondoftwo}%
\providecommand \href [0]{\begingroup \@sanitize@url \@href}%
\providecommand \@href[1]{\@@startlink{#1}\@@href}%
\providecommand \@@href[1]{\endgroup#1\@@endlink}%
\providecommand \@sanitize@url [0]{\catcode `\\12\catcode `\$12\catcode
  `\&12\catcode `\#12\catcode `\^12\catcode `\_12\catcode `\%12\relax}%
\providecommand \@@startlink[1]{}%
\providecommand \@@endlink[0]{}%
\providecommand \url  [0]{\begingroup\@sanitize@url \@url }%
\providecommand \@url [1]{\endgroup\@href {#1}{\urlprefix }}%
\providecommand \urlprefix  [0]{URL }%
\providecommand \Eprint [0]{\href }%
\providecommand \doibase [0]{http://dx.doi.org/}%
\providecommand \selectlanguage [0]{\@gobble}%
\providecommand \bibinfo  [0]{\@secondoftwo}%
\providecommand \bibfield  [0]{\@secondoftwo}%
\providecommand \translation [1]{[#1]}%
\providecommand \BibitemOpen [0]{}%
\providecommand \bibitemStop [0]{}%
\providecommand \bibitemNoStop [0]{.\EOS\space}%
\providecommand \EOS [0]{\spacefactor3000\relax}%
\providecommand \BibitemShut  [1]{\csname bibitem#1\endcsname}%
\let\auto@bib@innerbib\@empty
%</preamble>
\bibitem [{\citenamefont {Nakamura}\ \emph {et~al.}(1999)\citenamefont
  {Nakamura}, \citenamefont {Pashkin},\ and\ \citenamefont
  {Tsai}}]{Nakamura1999}%
  \BibitemOpen
  \bibfield  {author} {\bibinfo {author} {\bibfnamefont {Y.}~\bibnamefont
  {Nakamura}}, \bibinfo {author} {\bibfnamefont {Y.~A.}\ \bibnamefont
  {Pashkin}}, \ and\ \bibinfo {author} {\bibfnamefont {J.~S.}\ \bibnamefont
  {Tsai}},\ }\bibfield  {title} {\emph {\enquote {\bibinfo {title} {{Coherent
  control of macroscopic quantum states in a single-Cooper-pair box}},}\
  }}\href {\doibase 10.1038/19718} {\bibfield  {journal} {\bibinfo  {journal}
  {Nature}\ }\textbf {\bibinfo {volume} {398}},\ \bibinfo {pages} {786}
  (\bibinfo {year} {1999})}\BibitemShut {NoStop}%
\bibitem [{\citenamefont {{van der Wal}}\ \emph {et~al.}(1999)\citenamefont
  {{van der Wal}}, \citenamefont {Mooij}, \citenamefont {Orlando},
  \citenamefont {Levitov}, \citenamefont {Tian},\ and\ \citenamefont
  {Lloyd}}]{VanderWalCH1999}%
  \BibitemOpen
  \bibfield  {author} {\bibinfo {author} {\bibfnamefont {C.~H.}\ \bibnamefont
  {{van der Wal}}}, \bibinfo {author} {\bibfnamefont {J.~E.}\ \bibnamefont
  {Mooij}}, \bibinfo {author} {\bibfnamefont {T.~P.}\ \bibnamefont {Orlando}},
  \bibinfo {author} {\bibfnamefont {L.}~\bibnamefont {Levitov}}, \bibinfo
  {author} {\bibfnamefont {L.}~\bibnamefont {Tian}}, \ and\ \bibinfo {author}
  {\bibfnamefont {S.}~\bibnamefont {Lloyd}},\ }\bibfield  {title} {\emph
  {\enquote {\bibinfo {title} {{Josephson persistent-current qubit}},}\ }}\href
  {\doibase 10.1126/science.285.5430.1036} {\bibfield  {journal} {\bibinfo
  {journal} {Science}\ }\textbf {\bibinfo {volume} {285}},\ \bibinfo {pages}
  {1036} (\bibinfo {year} {1999})}\BibitemShut {NoStop}%
\bibitem [{\citenamefont {Vion}\ \emph {et~al.}(2002)\citenamefont {Vion},
  \citenamefont {Aassime}, \citenamefont {Cottet}, \citenamefont {Joyez},
  \citenamefont {Pothier}, \citenamefont {Urbina}, \citenamefont {Esteve},\
  and\ \citenamefont {Devoret}}]{Vion2002}%
  \BibitemOpen
  \bibfield  {author} {\bibinfo {author} {\bibfnamefont {D.}~\bibnamefont
  {Vion}}, \bibinfo {author} {\bibfnamefont {A.}~\bibnamefont {Aassime}},
  \bibinfo {author} {\bibfnamefont {A.}~\bibnamefont {Cottet}}, \bibinfo
  {author} {\bibfnamefont {P.}~\bibnamefont {Joyez}}, \bibinfo {author}
  {\bibfnamefont {H.}~\bibnamefont {Pothier}}, \bibinfo {author} {\bibfnamefont
  {C.}~\bibnamefont {Urbina}}, \bibinfo {author} {\bibfnamefont
  {D.}~\bibnamefont {Esteve}}, \ and\ \bibinfo {author} {\bibfnamefont {M.~H.}\
  \bibnamefont {Devoret}},\ }\bibfield  {title} {\emph {\enquote {\bibinfo
  {title} {{Manipulating the quantum state of an electrical circuit.}}}\
  }}\href {\doibase 10.1126/science.1069372} {\bibfield  {journal} {\bibinfo
  {journal} {Science}\ }\textbf {\bibinfo {volume} {296}},\ \bibinfo {pages}
  {886} (\bibinfo {year} {2002})}\BibitemShut {NoStop}%
\bibitem [{\citenamefont {Paik}\ \emph {et~al.}(2011)\citenamefont {Paik},
  \citenamefont {Schuster}, \citenamefont {Bishop}, \citenamefont {Kirchmair},
  \citenamefont {Catelani}, \citenamefont {Sears}, \citenamefont {Johnson},
  \citenamefont {Reagor}, \citenamefont {Frunzio}, \citenamefont {Glazman},
  \citenamefont {Girvin}, \citenamefont {Devoret},\ and\ \citenamefont
  {Schoelkopf}}]{Paik2011}%
  \BibitemOpen
  \bibfield  {author} {\bibinfo {author} {\bibfnamefont {H.}~\bibnamefont
  {Paik}}, \bibinfo {author} {\bibfnamefont {D.~I.}\ \bibnamefont {Schuster}},
  \bibinfo {author} {\bibfnamefont {L.~S.}\ \bibnamefont {Bishop}}, \bibinfo
  {author} {\bibfnamefont {G.}~\bibnamefont {Kirchmair}}, \bibinfo {author}
  {\bibfnamefont {G.}~\bibnamefont {Catelani}}, \bibinfo {author}
  {\bibfnamefont {A.~P.}\ \bibnamefont {Sears}}, \bibinfo {author}
  {\bibfnamefont {B.~R.}\ \bibnamefont {Johnson}}, \bibinfo {author}
  {\bibfnamefont {M.~J.}\ \bibnamefont {Reagor}}, \bibinfo {author}
  {\bibfnamefont {L.}~\bibnamefont {Frunzio}}, \bibinfo {author} {\bibfnamefont
  {L.~I.}\ \bibnamefont {Glazman}}, \bibinfo {author} {\bibfnamefont {S.~M.}\
  \bibnamefont {Girvin}}, \bibinfo {author} {\bibfnamefont {M.~H.}\
  \bibnamefont {Devoret}}, \ and\ \bibinfo {author} {\bibfnamefont {R.~J.}\
  \bibnamefont {Schoelkopf}},\ }\bibfield  {title} {\emph {\enquote {\bibinfo
  {title} {{Observation of high coherence in Josephson junction qubits measured
  in a three-dimensional circuit QED architecture}},}\ }}\href {\doibase
  10.1103/PhysRevLett.107.240501} {\bibfield  {journal} {\bibinfo  {journal}
  {Phys. Rev. Lett.}\ }\textbf {\bibinfo {volume} {107}},\ \bibinfo {pages}
  {240501} (\bibinfo {year} {2011})}\BibitemShut {NoStop}%
\bibitem [{\citenamefont {Barends}\ \emph {et~al.}(2013)\citenamefont
  {Barends}, \citenamefont {Kelly}, \citenamefont {Megrant}, \citenamefont
  {Sank}, \citenamefont {Jeffrey}, \citenamefont {Chen}, \citenamefont {Yin},
  \citenamefont {Chiaro}, \citenamefont {Mutus}, \citenamefont {Neill},
  \citenamefont {O'Malley}, \citenamefont {Roushan}, \citenamefont {Wenner},
  \citenamefont {White}, \citenamefont {Cleland},\ and\ \citenamefont
  {Martinis}}]{Barends2013}%
  \BibitemOpen
  \bibfield  {author} {\bibinfo {author} {\bibfnamefont {R.}~\bibnamefont
  {Barends}}, \bibinfo {author} {\bibfnamefont {J.}~\bibnamefont {Kelly}},
  \bibinfo {author} {\bibfnamefont {A.}~\bibnamefont {Megrant}}, \bibinfo
  {author} {\bibfnamefont {D.}~\bibnamefont {Sank}}, \bibinfo {author}
  {\bibfnamefont {E.}~\bibnamefont {Jeffrey}}, \bibinfo {author} {\bibfnamefont
  {Y.}~\bibnamefont {Chen}}, \bibinfo {author} {\bibfnamefont {Y.}~\bibnamefont
  {Yin}}, \bibinfo {author} {\bibfnamefont {B.}~\bibnamefont {Chiaro}},
  \bibinfo {author} {\bibfnamefont {J.}~\bibnamefont {Mutus}}, \bibinfo
  {author} {\bibfnamefont {C.}~\bibnamefont {Neill}}, \bibinfo {author}
  {\bibfnamefont {P.}~\bibnamefont {O'Malley}}, \bibinfo {author}
  {\bibfnamefont {P.}~\bibnamefont {Roushan}}, \bibinfo {author} {\bibfnamefont
  {J.}~\bibnamefont {Wenner}}, \bibinfo {author} {\bibfnamefont {T.~C.}\
  \bibnamefont {White}}, \bibinfo {author} {\bibfnamefont {A.~N.}\ \bibnamefont
  {Cleland}}, \ and\ \bibinfo {author} {\bibfnamefont {J.~M.}\ \bibnamefont
  {Martinis}},\ }\bibfield  {title} {\emph {\enquote {\bibinfo {title}
  {{Coherent Josephson qubit suitable for scalable quantum integrated
  circuits}},}\ }}\href {\doibase 10.1103/PhysRevLett.111.080502} {\bibfield
  {journal} {\bibinfo  {journal} {Phys. Rev. Lett.}\ }\textbf {\bibinfo
  {volume} {111}},\ \bibinfo {pages} {080502} (\bibinfo {year}
  {2013})}\BibitemShut {NoStop}%
\bibitem [{\citenamefont {Manucharyan}\ \emph {et~al.}(2009)\citenamefont
  {Manucharyan}, \citenamefont {Koch}, \citenamefont {Glazman},\ and\
  \citenamefont {Devoret}}]{Manucharyan2009}%
  \BibitemOpen
  \bibfield  {author} {\bibinfo {author} {\bibfnamefont {V.~E.}\ \bibnamefont
  {Manucharyan}}, \bibinfo {author} {\bibfnamefont {J.}~\bibnamefont {Koch}},
  \bibinfo {author} {\bibfnamefont {L.~I.}\ \bibnamefont {Glazman}}, \ and\
  \bibinfo {author} {\bibfnamefont {M.~H.}\ \bibnamefont {Devoret}},\
  }\bibfield  {title} {\emph {\enquote {\bibinfo {title} {{Fluxonium: Single
  Cooper-Pair Circuit Free of Charge Offsets}},}\ }}\href {\doibase
  10.1126/science.1175552} {\bibfield  {journal} {\bibinfo  {journal}
  {Science}\ }\textbf {\bibinfo {volume} {326}},\ \bibinfo {pages} {113}
  (\bibinfo {year} {2009})}\BibitemShut {NoStop}%
\bibitem [{\citenamefont {Masluk}\ \emph {et~al.}(2012)\citenamefont {Masluk},
  \citenamefont {Pop}, \citenamefont {Kamal}, \citenamefont {Minev},\ and\
  \citenamefont {Devoret}}]{Masluk2012}%
  \BibitemOpen
  \bibfield  {author} {\bibinfo {author} {\bibfnamefont {N.~A.}\ \bibnamefont
  {Masluk}}, \bibinfo {author} {\bibfnamefont {I.~M.}\ \bibnamefont {Pop}},
  \bibinfo {author} {\bibfnamefont {A.}~\bibnamefont {Kamal}}, \bibinfo
  {author} {\bibfnamefont {Z.~K.}\ \bibnamefont {Minev}}, \ and\ \bibinfo
  {author} {\bibfnamefont {M.~H.}\ \bibnamefont {Devoret}},\ }\bibfield
  {title} {\emph {\enquote {\bibinfo {title} {{Microwave characterization of
  Josephson junction arrays: Implementing a low loss superinductance}},}\
  }}\href {\doibase 10.1103/PhysRevLett.109.137002} {\bibfield  {journal}
  {\bibinfo  {journal} {Phys. Rev. Lett.}\ }\textbf {\bibinfo {volume} {109}},\
  \bibinfo {pages} {137002} (\bibinfo {year} {2012})}\BibitemShut {NoStop}%
\bibitem [{\citenamefont {Matveev}\ \emph {et~al.}(2002)\citenamefont
  {Matveev}, \citenamefont {Larkin},\ and\ \citenamefont
  {Glazman}}]{Matveev2002}%
  \BibitemOpen
  \bibfield  {author} {\bibinfo {author} {\bibfnamefont {K.~A.}\ \bibnamefont
  {Matveev}}, \bibinfo {author} {\bibfnamefont {A.~I.}\ \bibnamefont {Larkin}},
  \ and\ \bibinfo {author} {\bibfnamefont {L.~I.}\ \bibnamefont {Glazman}},\
  }\bibfield  {title} {\emph {\enquote {\bibinfo {title} {{Persistent current
  in superconducting nanorings.}}}\ }}\href {\doibase
  10.1103/PhysRevLett.89.096802} {\bibfield  {journal} {\bibinfo  {journal}
  {Phys. Rev. Lett.}\ }\textbf {\bibinfo {volume} {89}},\ \bibinfo {pages}
  {096802} (\bibinfo {year} {2002})}\BibitemShut {NoStop}%
\bibitem [{\citenamefont {Meier}\ \emph {et~al.}(2015)\citenamefont {Meier},
  \citenamefont {Brierley}, \citenamefont {Kou}, \citenamefont {Girvin},\ and\
  \citenamefont {Glazman}}]{Meier2015}%
  \BibitemOpen
  \bibfield  {author} {\bibinfo {author} {\bibfnamefont {H.}~\bibnamefont
  {Meier}}, \bibinfo {author} {\bibfnamefont {R.~T.}\ \bibnamefont {Brierley}},
  \bibinfo {author} {\bibfnamefont {A.}~\bibnamefont {Kou}}, \bibinfo {author}
  {\bibfnamefont {S.~M.}\ \bibnamefont {Girvin}}, \ and\ \bibinfo {author}
  {\bibfnamefont {L.~I.}\ \bibnamefont {Glazman}},\ }\bibfield  {title} {\emph
  {\enquote {\bibinfo {title} {{Signatures of quantum phase transitions in the
  dynamic response of fluxonium qubit chains}},}\ }}\href@noop {} {\bibfield
  {journal} {\bibinfo  {journal} {Phys. Rev. B}\ }\textbf {\bibinfo {volume}
  {92}},\ \bibinfo {pages} {064516} (\bibinfo {year} {2015})}\BibitemShut
  {NoStop}%
\bibitem [{sup()}]{superinductanceNote}%
  \BibitemOpen
  \href@noop {} {\bibinfo  {journal} {We expect the resonant frequency of the
  array mode to be $\sim12.5~$GHz, which is above the transition frequencies of
  interest for the molecule. We can then treat the array of junctions simply as
  a large inductance}\ }\BibitemShut {NoStop}%
\bibitem [{\citenamefont {Lecocq}\ \emph {et~al.}(2011)\citenamefont {Lecocq},
  \citenamefont {Pop}, \citenamefont {Peng}, \citenamefont {Matei},
  \citenamefont {Crozes}, \citenamefont {Fournier}, \citenamefont {Naud},
  \citenamefont {Guichard},\ and\ \citenamefont {Buisson}}]{Lecocq2011}%
  \BibitemOpen
\bibfield  {journal} {  }\bibfield  {author} {\bibinfo {author} {\bibfnamefont
  {F.}~\bibnamefont {Lecocq}}, \bibinfo {author} {\bibfnamefont {I.~M.}\
  \bibnamefont {Pop}}, \bibinfo {author} {\bibfnamefont {Z.}~\bibnamefont
  {Peng}}, \bibinfo {author} {\bibfnamefont {I.}~\bibnamefont {Matei}},
  \bibinfo {author} {\bibfnamefont {T.}~\bibnamefont {Crozes}}, \bibinfo
  {author} {\bibfnamefont {T.}~\bibnamefont {Fournier}}, \bibinfo {author}
  {\bibfnamefont {C.}~\bibnamefont {Naud}}, \bibinfo {author} {\bibfnamefont
  {W.}~\bibnamefont {Guichard}}, \ and\ \bibinfo {author} {\bibfnamefont
  {O.}~\bibnamefont {Buisson}},\ }\bibfield  {title} {\emph {\enquote {\bibinfo
  {title} {{Junction fabrication by shadow evaporation without a suspended
  bridge.}}}\ }}\href {\doibase 10.1088/0957-4484/22/31/315302} {\bibfield
  {journal} {\bibinfo  {journal} {Nanotechnology}\ }\textbf {\bibinfo {volume}
  {22}},\ \bibinfo {pages} {315302} (\bibinfo {year} {2011})}\BibitemShut
  {NoStop}%
\bibitem [{\citenamefont {Geerlings}()}]{KurtisThesis}%
  \BibitemOpen
  \bibfield  {author} {\bibinfo {author} {\bibfnamefont {K.~L.}\ \bibnamefont
  {Geerlings}},\ }\href@noop {} {\emph {\bibinfo {title} {Improving Coherence
  of Superconducting Qubits and Resonators}}}\ (\bibinfo  {publisher} {Yale
  University Press, 2013})\BibitemShut {NoStop}%
\bibitem [{\citenamefont {Blais}\ \emph {et~al.}(2004)\citenamefont {Blais},
  \citenamefont {Huang}, \citenamefont {Wallraff}, \citenamefont {Girvin},\
  and\ \citenamefont {Schoelkopf}}]{Blais2004}%
  \BibitemOpen
  \bibfield  {author} {\bibinfo {author} {\bibfnamefont {A.}~\bibnamefont
  {Blais}}, \bibinfo {author} {\bibfnamefont {R.~S.}\ \bibnamefont {Huang}},
  \bibinfo {author} {\bibfnamefont {A.}~\bibnamefont {Wallraff}}, \bibinfo
  {author} {\bibfnamefont {S.~M.}\ \bibnamefont {Girvin}}, \ and\ \bibinfo
  {author} {\bibfnamefont {R.~J.}\ \bibnamefont {Schoelkopf}},\ }\bibfield
  {title} {\emph {\enquote {\bibinfo {title} {{Cavity quantum electrodynamics
  for superconducting electrical circuits: An architecture for quantum
  computation}},}\ }}\href {\doibase 10.1103/PhysRevA.69.062320} {\bibfield
  {journal} {\bibinfo  {journal} {Phys. Rev. A}\ }\textbf {\bibinfo {volume}
  {69}},\ \bibinfo {pages} {062320} (\bibinfo {year} {2004})}\BibitemShut
  {NoStop}%
\bibitem [{\citenamefont {Gladchenko}\ \emph {et~al.}(2009)\citenamefont
  {Gladchenko}, \citenamefont {Olaya}, \citenamefont {Dupont-Ferrier},
  \citenamefont {Doucot}, \citenamefont {Ioffe},\ and\ \citenamefont
  {Gershenson}}]{Gladchenko2009}%
  \BibitemOpen
  \bibfield  {author} {\bibinfo {author} {\bibfnamefont {S.}~\bibnamefont
  {Gladchenko}}, \bibinfo {author} {\bibfnamefont {D.}~\bibnamefont {Olaya}},
  \bibinfo {author} {\bibfnamefont {E.}~\bibnamefont {Dupont-Ferrier}},
  \bibinfo {author} {\bibfnamefont {B.}~\bibnamefont {Doucot}}, \bibinfo
  {author} {\bibfnamefont {L.~B.}\ \bibnamefont {Ioffe}}, \ and\ \bibinfo
  {author} {\bibfnamefont {M.~E.}\ \bibnamefont {Gershenson}},\ }\bibfield
  {title} {\emph {\enquote {\bibinfo {title} {{Superconducting nanocircuits for
  topologically protected qubits}},}\ }}\href {\doibase 10.1038/NPHYS1151}
  {\bibfield  {journal} {\bibinfo  {journal} {Nat. Phys.}\ }\textbf {\bibinfo
  {volume} {5}},\ \bibinfo {pages} {48} (\bibinfo {year} {2009})}\BibitemShut
  {NoStop}%
\bibitem [{\citenamefont {Smith}\ \emph {et~al.}(2016)\citenamefont {Smith},
  \citenamefont {Kou}, \citenamefont {Vool}, \citenamefont {Pop}, \citenamefont
  {Frunzio}, \citenamefont {Schoelkopf},\ and\ \citenamefont
  {Devoret}}]{Smith2016}%
  \BibitemOpen
  \bibfield  {author} {\bibinfo {author} {\bibfnamefont {W.~C.}\ \bibnamefont
  {Smith}}, \bibinfo {author} {\bibfnamefont {A.}~\bibnamefont {Kou}}, \bibinfo
  {author} {\bibfnamefont {U.}~\bibnamefont {Vool}}, \bibinfo {author}
  {\bibfnamefont {I.~M.}\ \bibnamefont {Pop}}, \bibinfo {author} {\bibfnamefont
  {L.}~\bibnamefont {Frunzio}}, \bibinfo {author} {\bibfnamefont {R.~J.}\
  \bibnamefont {Schoelkopf}}, \ and\ \bibinfo {author} {\bibfnamefont {M.~H.}\
  \bibnamefont {Devoret}},\ }\bibfield  {title} {\emph {\enquote {\bibinfo
  {title} {{Quantization of inductively-shunted superconducting circuits}},}\
  }}\href {http://arxiv.org/abs/1602.01793} {\bibfield  {journal} {\bibinfo
  {journal} {Phys. Rev. B}\ }\textbf {\bibinfo {volume} {94}},\ \bibinfo
  {pages} {144507} (\bibinfo {year} {2016})}\BibitemShut {NoStop}%
\bibitem [{\citenamefont {Yoshihara}\ \emph {et~al.}(2006)\citenamefont
  {Yoshihara}, \citenamefont {Harrabi}, \citenamefont {Niskanen}, \citenamefont
  {Nakamura},\ and\ \citenamefont {Tsai}}]{Yoshihara2006}%
  \BibitemOpen
  \bibfield  {author} {\bibinfo {author} {\bibfnamefont {F.}~\bibnamefont
  {Yoshihara}}, \bibinfo {author} {\bibfnamefont {K.}~\bibnamefont {Harrabi}},
  \bibinfo {author} {\bibfnamefont {A.~O.}\ \bibnamefont {Niskanen}}, \bibinfo
  {author} {\bibfnamefont {Y.}~\bibnamefont {Nakamura}}, \ and\ \bibinfo
  {author} {\bibfnamefont {J.~S.}\ \bibnamefont {Tsai}},\ }\bibfield  {title}
  {\emph {\enquote {\bibinfo {title} {{Decoherence of flux qubits due to 1/f
  flux noise}},}\ }}\href {\doibase 10.1103/PhysRevLett.97.167001} {\bibfield
  {journal} {\bibinfo  {journal} {Phys. Rev. Lett.}\ }\textbf {\bibinfo
  {volume} {97}},\ \bibinfo {pages} {167001} (\bibinfo {year}
  {2006})}\BibitemShut {NoStop}%
\bibitem [{\citenamefont {Bylander}\ \emph {et~al.}(2011)\citenamefont
  {Bylander}, \citenamefont {Gustavsson}, \citenamefont {Yan}, \citenamefont
  {Yoshihara}, \citenamefont {Harrabi}, \citenamefont {Fitch}, \citenamefont
  {Cory}, \citenamefont {Nakamura}, \citenamefont {Tsai},\ and\ \citenamefont
  {Oliver}}]{Bylander2011}%
  \BibitemOpen
  \bibfield  {author} {\bibinfo {author} {\bibfnamefont {J.}~\bibnamefont
  {Bylander}}, \bibinfo {author} {\bibfnamefont {S.}~\bibnamefont
  {Gustavsson}}, \bibinfo {author} {\bibfnamefont {F.}~\bibnamefont {Yan}},
  \bibinfo {author} {\bibfnamefont {F.}~\bibnamefont {Yoshihara}}, \bibinfo
  {author} {\bibfnamefont {K.}~\bibnamefont {Harrabi}}, \bibinfo {author}
  {\bibfnamefont {G.}~\bibnamefont {Fitch}}, \bibinfo {author} {\bibfnamefont
  {D.~G.}\ \bibnamefont {Cory}}, \bibinfo {author} {\bibfnamefont
  {Y.}~\bibnamefont {Nakamura}}, \bibinfo {author} {\bibfnamefont {J.-S.}\
  \bibnamefont {Tsai}}, \ and\ \bibinfo {author} {\bibfnamefont {W.~D.}\
  \bibnamefont {Oliver}},\ }\bibfield  {title} {\emph {\enquote {\bibinfo
  {title} {{Dynamical decoupling and noise spectroscopy with a superconducting
  flux qubit - Supplement}},}\ }}\href {\doibase 10.1038/nphys1994} {\bibfield
  {journal} {\bibinfo  {journal} {Nat. Phys.}\ }\textbf {\bibinfo {volume}
  {7}},\ \bibinfo {pages} {21} (\bibinfo {year} {2011})}\BibitemShut {NoStop}%
\bibitem [{\citenamefont {Yoshihara}\ \emph {et~al.}(2010)\citenamefont
  {Yoshihara}, \citenamefont {Nakamura},\ and\ \citenamefont
  {Tsai}}]{Yoshihara2010}%
  \BibitemOpen
  \bibfield  {author} {\bibinfo {author} {\bibfnamefont {F.}~\bibnamefont
  {Yoshihara}}, \bibinfo {author} {\bibfnamefont {Y.}~\bibnamefont {Nakamura}},
  \ and\ \bibinfo {author} {\bibfnamefont {J.~S.}\ \bibnamefont {Tsai}},\
  }\bibfield  {title} {\emph {\enquote {\bibinfo {title} {{Correlated flux
  noise and decoherence in two inductively coupled flux qubits}},}\ }}\href
  {\doibase 10.1103/PhysRevB.81.132502} {\bibfield  {journal} {\bibinfo
  {journal} {Phys. Rev. B}\ }\textbf {\bibinfo {volume} {81}},\ \bibinfo
  {pages} {1} (\bibinfo {year} {2010})}\BibitemShut {NoStop}%
\bibitem [{\citenamefont {Koch}\ \emph {et~al.}(2007)\citenamefont {Koch},
  \citenamefont {Divincenzo},\ and\ \citenamefont {Clarke}}]{Koch2007}%
  \BibitemOpen
  \bibfield  {author} {\bibinfo {author} {\bibfnamefont {R.~H.}\ \bibnamefont
  {Koch}}, \bibinfo {author} {\bibfnamefont {D.~P.}\ \bibnamefont
  {Divincenzo}}, \ and\ \bibinfo {author} {\bibfnamefont {J.}~\bibnamefont
  {Clarke}},\ }\bibfield  {title} {\emph {\enquote {\bibinfo {title} {{Model
  for 1/f flux noise in SQUIDs and qubits}},}\ }}\href {\doibase
  10.1103/PhysRevLett.98.267003} {\bibfield  {journal} {\bibinfo  {journal}
  {Phys. Rev. Lett.}\ }\textbf {\bibinfo {volume} {98}},\ \bibinfo {pages}
  {267003} (\bibinfo {year} {2007})}\BibitemShut {NoStop}%
\bibitem [{\citenamefont {Faoro}\ and\ \citenamefont
  {Ioffe}(2008)}]{Faoro2008}%
  \BibitemOpen
  \bibfield  {author} {\bibinfo {author} {\bibfnamefont {L.}~\bibnamefont
  {Faoro}}\ and\ \bibinfo {author} {\bibfnamefont {L.~B.}\ \bibnamefont
  {Ioffe}},\ }\bibfield  {title} {\emph {\enquote {\bibinfo {title}
  {{Microscopic origin of low-frequency flux noise in Josephson circuits}},}\
  }}\href {\doibase 10.1103/PhysRevLett.100.227005} {\bibfield  {journal}
  {\bibinfo  {journal} {Phys. Rev. Lett.}\ }\textbf {\bibinfo {volume} {100}},\
  \bibinfo {pages} {227005} (\bibinfo {year} {2008})}\BibitemShut {NoStop}%
\bibitem [{\citenamefont {Kumar}\ \emph {et~al.}(2016)\citenamefont {Kumar},
  \citenamefont {Sendelbach}, \citenamefont {Beck}, \citenamefont {Freeland},
  \citenamefont {Wang}, \citenamefont {Wang}, \citenamefont {Yu}, \citenamefont
  {Wu}, \citenamefont {Pappas},\ and\ \citenamefont {McDermott}}]{Kumar2016}%
  \BibitemOpen
  \bibfield  {author} {\bibinfo {author} {\bibfnamefont {P.}~\bibnamefont
  {Kumar}}, \bibinfo {author} {\bibfnamefont {S.}~\bibnamefont {Sendelbach}},
  \bibinfo {author} {\bibfnamefont {M.~A.}\ \bibnamefont {Beck}}, \bibinfo
  {author} {\bibfnamefont {J.~W.}\ \bibnamefont {Freeland}}, \bibinfo {author}
  {\bibfnamefont {Z.}~\bibnamefont {Wang}}, \bibinfo {author} {\bibfnamefont
  {H.}~\bibnamefont {Wang}}, \bibinfo {author} {\bibfnamefont {C.~C.}\
  \bibnamefont {Yu}}, \bibinfo {author} {\bibfnamefont {R.~Q.}\ \bibnamefont
  {Wu}}, \bibinfo {author} {\bibfnamefont {D.~P.}\ \bibnamefont {Pappas}}, \
  and\ \bibinfo {author} {\bibfnamefont {R.}~\bibnamefont {McDermott}},\
  }\bibfield  {title} {\emph {\enquote {\bibinfo {title} {{Origin and Reduction
  of 1/f Magnetic Flux Noise in Superconducting Devices}},}\ }}\href
  {http://arxiv.org/abs/1604.00877} {\bibfield  {journal} {\bibinfo  {journal}
  {Phys. Rev. Applied}\ }\textbf {\bibinfo {volume} {6}},\ \bibinfo {pages}
  {041001} (\bibinfo {year} {2016})}\BibitemShut {NoStop}%
\bibitem [{\citenamefont {Ioffe}\ \emph {et~al.}(2002)\citenamefont {Ioffe},
  \citenamefont {Feigel'man}, \citenamefont {Ioselevich}, \citenamefont
  {Ivanov}, \citenamefont {Troyer},\ and\ \citenamefont {Blatter}}]{Ioffe2002}%
  \BibitemOpen
  \bibfield  {author} {\bibinfo {author} {\bibfnamefont {L.~B.}\ \bibnamefont
  {Ioffe}}, \bibinfo {author} {\bibfnamefont {M.~V.}\ \bibnamefont
  {Feigel'man}}, \bibinfo {author} {\bibfnamefont {A.}~\bibnamefont
  {Ioselevich}}, \bibinfo {author} {\bibfnamefont {D.}~\bibnamefont {Ivanov}},
  \bibinfo {author} {\bibfnamefont {M.}~\bibnamefont {Troyer}}, \ and\ \bibinfo
  {author} {\bibfnamefont {G.}~\bibnamefont {Blatter}},\ }\bibfield  {title}
  {\emph {\enquote {\bibinfo {title} {{Topologically protected quantum bits
  using Josephson junction arrays}},}\ }}\href {\doibase 10.1038/415503a}
  {\bibfield  {journal} {\bibinfo  {journal} {Nature}\ }\textbf {\bibinfo
  {volume} {415}},\ \bibinfo {pages} {503} (\bibinfo {year}
  {2002})}\BibitemShut {NoStop}%
\bibitem [{\citenamefont {Brooks}\ \emph {et~al.}(2013)\citenamefont {Brooks},
  \citenamefont {Kitaev},\ and\ \citenamefont {Preskill}}]{Brooks2013}%
  \BibitemOpen
  \bibfield  {author} {\bibinfo {author} {\bibfnamefont {P.}~\bibnamefont
  {Brooks}}, \bibinfo {author} {\bibfnamefont {A.}~\bibnamefont {Kitaev}}, \
  and\ \bibinfo {author} {\bibfnamefont {J.}~\bibnamefont {Preskill}},\
  }\bibfield  {title} {\emph {\enquote {\bibinfo {title} {{Protected gates for
  superconducting qubits}},}\ }}\href {\doibase 10.1103/PhysRevA.87.052306}
  {\bibfield  {journal} {\bibinfo  {journal} {Phys. Rev. A}\ }\textbf {\bibinfo
  {volume} {87}},\ \bibinfo {pages} {052306} (\bibinfo {year}
  {2013})}\BibitemShut {NoStop}%
\bibitem [{\citenamefont {Ithier}\ \emph {et~al.}(2005)\citenamefont {Ithier},
  \citenamefont {Collin}, \citenamefont {Joyez}, \citenamefont {Meeson},
  \citenamefont {Vion}, \citenamefont {Esteve}, \citenamefont {Chiarello},
  \citenamefont {Shnirman}, \citenamefont {Makhlin}, \citenamefont {Schriefl},\
  and\ \citenamefont {Schon}}]{Ithier2005}%
  \BibitemOpen
  \bibfield  {author} {\bibinfo {author} {\bibfnamefont {G.}~\bibnamefont
  {Ithier}}, \bibinfo {author} {\bibfnamefont {E.}~\bibnamefont {Collin}},
  \bibinfo {author} {\bibfnamefont {P.}~\bibnamefont {Joyez}}, \bibinfo
  {author} {\bibfnamefont {P.~J.}\ \bibnamefont {Meeson}}, \bibinfo {author}
  {\bibfnamefont {D.}~\bibnamefont {Vion}}, \bibinfo {author} {\bibfnamefont
  {D.}~\bibnamefont {Esteve}}, \bibinfo {author} {\bibfnamefont
  {F.}~\bibnamefont {Chiarello}}, \bibinfo {author} {\bibfnamefont
  {A.}~\bibnamefont {Shnirman}}, \bibinfo {author} {\bibfnamefont
  {Y.}~\bibnamefont {Makhlin}}, \bibinfo {author} {\bibfnamefont
  {J.}~\bibnamefont {Schriefl}}, \ and\ \bibinfo {author} {\bibfnamefont
  {G.}~\bibnamefont {Schon}},\ }\bibfield  {title} {\emph {\enquote {\bibinfo
  {title} {{Decoherence in a superconducting quantum bit circuit}},}\
  }}\href@noop {} {\bibfield  {journal} {\bibinfo  {journal} {Phys. Rev. B}\
  }\textbf {\bibinfo {volume} {72}},\ \bibinfo {pages} {134519} (\bibinfo
  {year} {2005})}\BibitemShut {NoStop}%
\bibitem [{\citenamefont {Van~Harlingen}\ \emph {et~al.}(2004)\citenamefont
  {Van~Harlingen}, \citenamefont {Robertson}, \citenamefont {Plourde},
  \citenamefont {Reichardt}, \citenamefont {Crane},\ and\ \citenamefont
  {John}}]{Harlingen2004}%
  \BibitemOpen
  \bibfield  {author} {\bibinfo {author} {\bibfnamefont {D.~J.}\ \bibnamefont
  {Van~Harlingen}}, \bibinfo {author} {\bibfnamefont {T.~L.}\ \bibnamefont
  {Robertson}}, \bibinfo {author} {\bibfnamefont {B.~L.~T.}\ \bibnamefont
  {Plourde}}, \bibinfo {author} {\bibfnamefont {P.~A.}\ \bibnamefont
  {Reichardt}}, \bibinfo {author} {\bibfnamefont {T.~A.}\ \bibnamefont
  {Crane}}, \ and\ \bibinfo {author} {\bibfnamefont {C.}~\bibnamefont {John}},\
  }\bibfield  {title} {\emph {\enquote {\bibinfo {title} {{Decoherence in
  Josephson-junction qubits due to critical-current fluctuations}},}\
  }}\href@noop {} {\bibfield  {journal} {\bibinfo  {journal} {Phys. Rev. B}\
  }\textbf {\bibinfo {volume} {70}},\ \bibinfo {pages} {064517} (\bibinfo
  {year} {2004})}\BibitemShut {NoStop}%
\bibitem [{\citenamefont {Gambetta}\ \emph {et~al.}(2006)\citenamefont
  {Gambetta}, \citenamefont {Blais}, \citenamefont {Schuster}, \citenamefont
  {Wallraff}, \citenamefont {Frunzio}, \citenamefont {Majer}, \citenamefont
  {Devoret}, \citenamefont {Girvin},\ and\ \citenamefont
  {Schoelkopf}}]{Gambetta2006}%
  \BibitemOpen
  \bibfield  {author} {\bibinfo {author} {\bibfnamefont {J.}~\bibnamefont
  {Gambetta}}, \bibinfo {author} {\bibfnamefont {A.}~\bibnamefont {Blais}},
  \bibinfo {author} {\bibfnamefont {D.~I.}\ \bibnamefont {Schuster}}, \bibinfo
  {author} {\bibfnamefont {A.}~\bibnamefont {Wallraff}}, \bibinfo {author}
  {\bibfnamefont {L.}~\bibnamefont {Frunzio}}, \bibinfo {author} {\bibfnamefont
  {J.}~\bibnamefont {Majer}}, \bibinfo {author} {\bibfnamefont {M.~H.}\
  \bibnamefont {Devoret}}, \bibinfo {author} {\bibfnamefont {S.~M.}\
  \bibnamefont {Girvin}}, \ and\ \bibinfo {author} {\bibfnamefont {R.~J.}\
  \bibnamefont {Schoelkopf}},\ }\bibfield  {title} {\emph {\enquote {\bibinfo
  {title} {{Qubit-photon interactions in a cavity: Measurement-induced
  dephasing and number splitting}},}\ }}\href {\doibase
  10.1103/PhysRevA.74.042318} {\bibfield  {journal} {\bibinfo  {journal} {Phys.
  Rev. A}\ }\textbf {\bibinfo {volume} {74}},\ \bibinfo {pages} {042318}
  (\bibinfo {year} {2006})}\BibitemShut {NoStop}%
\bibitem [{\citenamefont {Clerk}\ and\ \citenamefont
  {Utami}(2007)}]{Clerk2007}%
  \BibitemOpen
  \bibfield  {author} {\bibinfo {author} {\bibfnamefont {A.~A.}\ \bibnamefont
  {Clerk}}\ and\ \bibinfo {author} {\bibfnamefont {D.~W.}\ \bibnamefont
  {Utami}},\ }\bibfield  {title} {\emph {\enquote {\bibinfo {title} {{Using a
  qubit to measure photon-number statistics of a driven thermal oscillator}},}\
  }}\href {\doibase 10.1103/PhysRevA.75.042302} {\bibfield  {journal} {\bibinfo
   {journal} {Phys. Rev. A}\ }\textbf {\bibinfo {volume} {75}},\ \bibinfo
  {pages} {042302} (\bibinfo {year} {2007})}\BibitemShut {NoStop}%
\bibitem [{\citenamefont {Yan}\ \emph {et~al.}(2015)\citenamefont {Yan},
  \citenamefont {Gustavsson}, \citenamefont {Kamal}, \citenamefont {Birenbaum},
  \citenamefont {Sears}, \citenamefont {Hover}, \citenamefont {Gudmundsen},
  \citenamefont {Yoder}, \citenamefont {Orlando}, \citenamefont {Clarke},
  \citenamefont {Kerman},\ and\ \citenamefont {Oliver}}]{Yan2015}%
  \BibitemOpen
  \bibfield  {author} {\bibinfo {author} {\bibfnamefont {F.}~\bibnamefont
  {Yan}}, \bibinfo {author} {\bibfnamefont {S.}~\bibnamefont {Gustavsson}},
  \bibinfo {author} {\bibfnamefont {A.}~\bibnamefont {Kamal}}, \bibinfo
  {author} {\bibfnamefont {J.}~\bibnamefont {Birenbaum}}, \bibinfo {author}
  {\bibfnamefont {a.~P.}\ \bibnamefont {Sears}}, \bibinfo {author}
  {\bibfnamefont {D.}~\bibnamefont {Hover}}, \bibinfo {author} {\bibfnamefont
  {G.~S. T.~J.}\ \bibnamefont {Gudmundsen}}, \bibinfo {author} {\bibfnamefont
  {J.~L.}\ \bibnamefont {Yoder}}, \bibinfo {author} {\bibfnamefont {T.~P.}\
  \bibnamefont {Orlando}}, \bibinfo {author} {\bibfnamefont {J.}~\bibnamefont
  {Clarke}}, \bibinfo {author} {\bibfnamefont {A.~J.}\ \bibnamefont {Kerman}},
  \ and\ \bibinfo {author} {\bibfnamefont {W.~D.}\ \bibnamefont {Oliver}},\
  }\bibfield  {title} {\emph {\enquote {\bibinfo {title} {{The Flux Qubit
  Revisited}},}\ }}\href {http://arxiv.org/abs/1508.06299} {\bibfield
  {journal} {\bibinfo  {journal} {arXiv:1508.06299}\ } (\bibinfo {year}
  {2015})}\BibitemShut {NoStop}%
\bibitem [{\citenamefont {Kou}()}]{unpubAkou}%
  \BibitemOpen
  \bibfield  {author} {\bibinfo {author} {\bibfnamefont {A.}~\bibnamefont
  {Kou}},\ }\href@noop {} {\bibinfo  {journal} {in preparation}\ }\BibitemShut
  {NoStop}%
\bibitem [{\citenamefont {Pop}\ \emph {et~al.}(2014)\citenamefont {Pop},
  \citenamefont {Geerlings}, \citenamefont {Catelani}, \citenamefont
  {Schoelkopf}, \citenamefont {Glazman},\ and\ \citenamefont
  {Devoret}}]{Pop2014}%
  \BibitemOpen
\bibfield  {journal} {  }\bibfield  {author} {\bibinfo {author} {\bibfnamefont
  {I.~M.}\ \bibnamefont {Pop}}, \bibinfo {author} {\bibfnamefont
  {K.}~\bibnamefont {Geerlings}}, \bibinfo {author} {\bibfnamefont
  {G.}~\bibnamefont {Catelani}}, \bibinfo {author} {\bibfnamefont {R.~J.}\
  \bibnamefont {Schoelkopf}}, \bibinfo {author} {\bibfnamefont {L.~I.}\
  \bibnamefont {Glazman}}, \ and\ \bibinfo {author} {\bibfnamefont {M.~H.}\
  \bibnamefont {Devoret}},\ }\bibfield  {title} {\emph {\enquote {\bibinfo
  {title} {{Coherent suppression of electromagnetic dissipation due to
  superconducting quasiparticles.}}}\ }}\href {\doibase 10.1038/nature13017}
  {\bibfield  {journal} {\bibinfo  {journal} {Nature}\ }\textbf {\bibinfo
  {volume} {508}},\ \bibinfo {pages} {369} (\bibinfo {year}
  {2014})}\BibitemShut {NoStop}%
\bibitem [{\citenamefont {Ludlow}\ \emph {et~al.}(2006)\citenamefont {Ludlow},
  \citenamefont {Boyd}, \citenamefont {Zelevinsky}, \citenamefont {Foreman},
  \citenamefont {Blatt}, \citenamefont {Notcutt}, \citenamefont {Ido},\ and\
  \citenamefont {Ye}}]{Ludlow2006}%
  \BibitemOpen
  \bibfield  {author} {\bibinfo {author} {\bibfnamefont {A.~D.}\ \bibnamefont
  {Ludlow}}, \bibinfo {author} {\bibfnamefont {M.~M.}\ \bibnamefont {Boyd}},
  \bibinfo {author} {\bibfnamefont {T.}~\bibnamefont {Zelevinsky}}, \bibinfo
  {author} {\bibfnamefont {S.~M.}\ \bibnamefont {Foreman}}, \bibinfo {author}
  {\bibfnamefont {S.}~\bibnamefont {Blatt}}, \bibinfo {author} {\bibfnamefont
  {M.}~\bibnamefont {Notcutt}}, \bibinfo {author} {\bibfnamefont
  {T.}~\bibnamefont {Ido}}, \ and\ \bibinfo {author} {\bibfnamefont
  {J.}~\bibnamefont {Ye}},\ }\bibfield  {title} {\emph {\enquote {\bibinfo
  {title} {{Systematic study of the Sr87 clock transition in an optical
  lattice}},}\ }}\href {\doibase 10.1103/PhysRevLett.96.033003} {\bibfield
  {journal} {\bibinfo  {journal} {Phys. Rev. Lett.}\ }\textbf {\bibinfo
  {volume} {96}},\ \bibinfo {pages} {033003} (\bibinfo {year}
  {2006})}\BibitemShut {NoStop}%
\bibitem [{\citenamefont {Langer}\ \emph {et~al.}(2005)\citenamefont {Langer},
  \citenamefont {Ozeri}, \citenamefont {Jost}, \citenamefont {Chiaverini},
  \citenamefont {Demarco}, \citenamefont {Ben-Kish}, \citenamefont {Blakestad},
  \citenamefont {Britton}, \citenamefont {Hume}, \citenamefont {Itano},
  \citenamefont {Leibfried}, \citenamefont {Reichle}, \citenamefont
  {Rosenband}, \citenamefont {Schaetz}, \citenamefont {Schmidt},\ and\
  \citenamefont {Wineland}}]{Langer2005}%
  \BibitemOpen
  \bibfield  {author} {\bibinfo {author} {\bibfnamefont {C.}~\bibnamefont
  {Langer}}, \bibinfo {author} {\bibfnamefont {R.}~\bibnamefont {Ozeri}},
  \bibinfo {author} {\bibfnamefont {J.~D.}\ \bibnamefont {Jost}}, \bibinfo
  {author} {\bibfnamefont {J.}~\bibnamefont {Chiaverini}}, \bibinfo {author}
  {\bibfnamefont {B.}~\bibnamefont {Demarco}}, \bibinfo {author} {\bibfnamefont
  {A.}~\bibnamefont {Ben-Kish}}, \bibinfo {author} {\bibfnamefont {R.~B.}\
  \bibnamefont {Blakestad}}, \bibinfo {author} {\bibfnamefont {J.}~\bibnamefont
  {Britton}}, \bibinfo {author} {\bibfnamefont {D.~B.}\ \bibnamefont {Hume}},
  \bibinfo {author} {\bibfnamefont {W.~M.}\ \bibnamefont {Itano}}, \bibinfo
  {author} {\bibfnamefont {D.}~\bibnamefont {Leibfried}}, \bibinfo {author}
  {\bibfnamefont {R.}~\bibnamefont {Reichle}}, \bibinfo {author} {\bibfnamefont
  {T.}~\bibnamefont {Rosenband}}, \bibinfo {author} {\bibfnamefont
  {T.}~\bibnamefont {Schaetz}}, \bibinfo {author} {\bibfnamefont {P.~O.}\
  \bibnamefont {Schmidt}}, \ and\ \bibinfo {author} {\bibfnamefont {D.~J.}\
  \bibnamefont {Wineland}},\ }\bibfield  {title} {\emph {\enquote {\bibinfo
  {title} {{Long-lived qubit memory using atomic ions}},}\ }}\href {\doibase
  10.1103/PhysRevLett.95.060502} {\bibfield  {journal} {\bibinfo  {journal}
  {Phys. Rev. Lett.}\ }\textbf {\bibinfo {volume} {95}},\ \bibinfo {pages}
  {060502} (\bibinfo {year} {2005})}\BibitemShut {NoStop}%
\bibitem [{\citenamefont {Bergmann}\ \emph {et~al.}(1998)\citenamefont
  {Bergmann}, \citenamefont {Theuer},\ and\ \citenamefont
  {Shore}}]{STIRAPreview}%
  \BibitemOpen
  \bibfield  {author} {\bibinfo {author} {\bibfnamefont {K.}~\bibnamefont
  {Bergmann}}, \bibinfo {author} {\bibfnamefont {H.}~\bibnamefont {Theuer}}, \
  and\ \bibinfo {author} {\bibfnamefont {B.~W.}\ \bibnamefont {Shore}},\
  }\bibfield  {title} {\emph {\enquote {\bibinfo {title} {{Coherent population
  transfer among quantum states of atoms and molecules}},}\ }}\href@noop {}
  {\bibfield  {journal} {\bibinfo  {journal} {Rev. Mod. Phys.}\ }\textbf
  {\bibinfo {volume} {70}},\ \bibinfo {pages} {1003} (\bibinfo {year}
  {1998})}\BibitemShut {NoStop}%
\end{thebibliography}%

\end{document}